%% file: main.tex
\pdfoutput=1
\documentclass[conference,compsoc]{IEEEtran}

\usepackage[T1]{fontenc}
\usepackage{amsmath,amssymb}
\makeatletter\@ifundefined{proof}{}{}\makeatother %
\usepackage{amsthm}
\usepackage{stmaryrd}            %
\usepackage{mathpartir}          %
\usepackage{graphicx}
\usepackage{booktabs,multirow,makecell,colortbl}
\usepackage{xcolor}
\usepackage{listings}
\usepackage{xspace}
\usepackage{pifont}
\usepackage{enumitem}
\usepackage[most]{tcolorbox}
\usepackage{cite}

\usepackage[hyphens]{url}
\usepackage{hyperref}
\usepackage[capitalise,nameinlink]{cleveref}
\hypersetup{colorlinks=true, linkcolor=[rgb]{0.10,0.20,0.50}, citecolor=[rgb]{0.10,0.40,0.20}, urlcolor=[rgb]{0.25,0.25,0.55}}

\input{macros}

\begin{document}
\raggedbottom   %
\makeatletter\@secpenalty=0\makeatother   %

\title{Suspected Until Inspected: Neuro-Symbolic Detection of\\Broken Access Control in Server-Side Web Applications}
\author{\IEEEauthorblockN{Chengyao Yang\IEEEauthorrefmark{1},
Zifeng Kang\IEEEauthorrefmark{1},
Dongqi Han, and
Shihong Zou\IEEEauthorrefmark{2}}
\IEEEauthorblockA{Beijing University of Posts and Telecommunications\\
\{cyyang, zifengkang, handongqi, zoush\}@bupt.edu.cn\\[2pt]
\IEEEauthorrefmark{1}Equal contribution.\quad \IEEEauthorrefmark{2}Corresponding author.}}
\maketitle

\input{sections/abstract}
\input{sections/intro}
\input{sections/problem}
\input{sections/overview}
\input{sections/design}
\input{sections/implementation}
\input{sections/evaluation}
\input{sections/discussion}
\input{sections/related}
\input{sections/conclusion}

\IfFileExists{IEEEtran.bst}{\bibliographystyle{IEEEtran}}{\bibliographystyle{ieeetr}}
\bibliography{refs}

\appendices
\crefalias{section}{appendix}
\input{sections/appendix}

\end{document}

%% file: macros.tex
\newcommand{\sys}{FLINT\xspace}

\newcommand{\cmark}{\ding{51}}

\newcommand{\code}[1]{\texttt{#1}}                 %

\newcommand{\tcode}[1]{\texttt{\footnotesize #1}}  %
\newcommand{\icode}[1]{\texttt{\scriptsize #1}}    %
\newcommand{\rn}[1]{\textsc{#1}}                   %
\newcommand{\kind}[1]{\texttt{#1}}                 %
\newcommand{\why}[1]{\texttt{#1}}                  %
\newcommand{\tier}[1]{#1}
\newcommand{\para}[1]{\vspace{1.5pt plus 1pt minus 0.5pt}\noindent\textbf{#1}\hspace{0.45em}}

\newcommand{\etal}{et al.\xspace}
\newcommand{\RQ}[2]{\textbf{RQ#1} [#2]}           %
\newcommand{\RQtag}[1]{[RQ#1]}

\newcommand{\orig}{\mathit{orig}}
\newcommand{\Rc}{R_{\mathit{me}}}                  %
\newcommand{\Client}{\mathit{Client}}

\newcommand{\Sel}{\mathit{Sel}}
\newcommand{\Row}{\mathit{Row}}
\newcommand{\lin}{\rightsquigarrow}                %
\newcommand{\rootP}{\ensuremath{\mathtt{P}}}
\newcommand{\rootS}{\ensuremath{\mathtt{S}}}
\newcommand{\rootG}{\ensuremath{\mathtt{G}}}
\newcommand{\rootL}{\ensuremath{\mathtt{L}}}
\newcommand{\rootU}{\ensuremath{\mathtt{U}}}
\newcommand{\dstep}[1]{{\triangleright}\mathit{#1}}         %
\newcommand{\kstep}[1]{{\blacktriangleright}\mathit{#1}}    %
\newcommand{\ext}[1]{{#1}^{\uparrow}}              %
\newcommand{\Obj}{\mathit{Obj}}
\newcommand{\ops}{\mathit{ops}}
\newcommand{\lvl}{\mathit{lvl}}
\newcommand{\st}{\mathit{st}}
\newcommand{\VALID}{\textsc{Valid}\xspace}
\newcommand{\REFUTED}{\textsc{Refuted}\xspace}
\newcommand{\UNVERIFIED}{\textsc{Unverified}\xspace}
\newcommand{\mValid}{\textsc{Valid}}
\newcommand{\mRefuted}{\textsc{Refuted}}
\newcommand{\mUnverified}{\textsc{Unverified}}
\newcommand{\need}{\mathit{need}}
\newcommand{\auth}{\mathit{auth}}
\newcommand{\contrib}{\mathit{con}}

\newcommand{\keys}{\mathit{key}}
\newcommand{\Links}{\mathit{Lk}}
\newcommand{\sem}[1]{\llbracket #1 \rrbracket}
\newcommand{\Tc}{\mathcal{T}}
\newcommand{\Fc}{\mathcal{F}}
\newcommand{\neutral}{\varepsilon}
\newcommand{\Cleared}{\textsc{Cleared}\xspace}
\newcommand{\Finding}{\textsc{Finding}\xspace}
\newcommand{\Lead}{\textsc{Lead}\xspace}
\newcommand{\KeyCovered}{\mathit{KeyCovered}}
\newcommand{\Broken}{\mathit{BrokenGuard}}

\theoremstyle{definition}
\newtheorem{definition}{Definition}
\newtheorem{theorem}{Theorem}
\newtheorem{lemma}{Lemma}
\newtheorem{example}{Example}
\crefname{figure}{Figure}{Figures}
\Crefname{figure}{Figure}{Figures}
\crefname{table}{Table}{Tables}
\Crefname{table}{Table}{Tables}
\crefname{section}{Section}{Sections}
\Crefname{section}{Section}{Sections}
\crefname{appendix}{Appendix}{Appendices}
\Crefname{appendix}{Appendix}{Appendices}
\crefname{equation}{Eq.}{Eqs.}
\crefname{lstlisting}{Listing}{Listings}
\Crefname{lstlisting}{Listing}{Listings}
\crefname{theorem}{Theorem}{Theorems}
\crefname{lemma}{Lemma}{Lemmas}
\crefname{definition}{Definition}{Definitions}
\crefname{example}{Example}{Examples}

\definecolor{ink}{HTML}{000000}
\definecolor{cmtgray}{HTML}{000000}   %
\definecolor{kwblue}{HTML}{1F5FA8}
\definecolor{strgreen}{HTML}{2E7D4F}
\definecolor{cmtmagenta}{HTML}{A23BB0}
\definecolor{neural}{HTML}{2E7D32}      %
\definecolor{neuralbg}{HTML}{EAF4EA}
\definecolor{symbolic}{HTML}{000000}    %
\definecolor{bugred}{HTML}{C0392B}
\definecolor{hlred}{HTML}{F9D9D3}
\definecolor{hlblue}{HTML}{DCE4FD}
\definecolor{hlgreen}{HTML}{D5EEDF}
\definecolor{hlyellow}{HTML}{FFF0A8}

\lstdefinelanguage{TypeScript}{morekeywords={async,await,const,let,var,if,else,throw,new,return,function,export,import,class,for,of,in,try,catch,this},
  sensitive=true, morecomment=[l]{//}, morecomment=[s]{/*}{*/}, morestring=[b]', morestring=[b]"}
\newcommand{\hlspan}[2]{{\setlength{\fboxsep}{0.6pt}\colorbox{#1}{\strut\ttfamily\fontsize{6.7}{7.9}\selectfont #2}}}
\makeatletter
\providecommand{\ext@lstlisting}{lol}
\AtBeginDocument{\def\fnum@lstlisting{\lstlistingname~\thelstlisting}}
\newcommand{\listingcaption}{\def\@captype{lstlisting}\caption}   %
\makeatother
\newcommand{\lcmt}[1]{{\color{cmtmagenta}\ttfamily\fontsize{6.7}{7.9}\selectfont /*#1*/}}

\newtcolorbox{promptbox}[1]{enhanced, breakable=false, sharp corners, colback=neuralbg, colframe=neural, boxrule=0.5pt,
  left=3pt, right=3pt, top=1.5pt, bottom=1.5pt, fonttitle=\footnotesize\bfseries, coltitle=white,
  colbacktitle=neural, title={#1}, fontupper=\scriptsize, toptitle=0.5pt, bottomtitle=0.5pt}
\newtcolorbox{answerbox}[1]{enhanced, breakable=false, sharp corners, colback=white, colframe=neural, boxrule=0.5pt,
  left=3pt, right=3pt, top=1.5pt, bottom=1.5pt, fonttitle=\footnotesize\bfseries, coltitle=white,
  colbacktitle=neural, title={#1}, fontupper=\scriptsize, toptitle=0.5pt, bottomtitle=0.5pt}
\lstdefinelanguage{json}{morestring=[b]", stringstyle=\color{ink}, literate={:}{{\textcolor{cmtgray}{:}}}1}

\setlist{leftmargin=1.1em, itemsep=1pt, parsep=0pt, topsep=2pt}

%% file: sections/abstract.tex
\begin{abstract}
Broken access control (BAC) has topped the OWASP Top 10 since 2021. Unfortunately, a decade of detectors still leaves a gap on every front. Static analyzers are tied to one language or framework, dynamic testers cover only the endpoints they manage to exercise, and LLM-based auditors leave the final verdict to an error-prone model.

In this paper, we design and implement FLINT, a neuro-symbolic framework that detects BAC in server-side web applications without language- or framework-specific rules. Our key insight is to separate the proposal from the verdict---an LLM may flag any data access as needing a guard, but only a symbolically verified check can clear it. The neural component enumerates every endpoint and distills each, via demand-driven code slicing, into an authorization certificate whose claims are all grounded in cited code. The symbolic component verifies each claimed check through origin tracing and a lattice-based interprocedural dataflow analysis, so that no access is cleared on the model's word.

When applied to popular real-world open-source applications, including Gitea, Grafana, and Strapi, FLINT detected 112 zero-day, exploitable BAC vulnerabilities, 30 of them in projects with over 20,000 GitHub stars. We have responsibly disclosed all of them to the developers. Mattermost, for example, has already confirmed our report and is assigning a CVE identifier.
\end{abstract}

%% file: sections/intro.tex
\section{Introduction}
\label{sec:intro}

Broken access control (BAC) has been the top risk of the OWASP Top~10 since 2021~\cite{owasp2021,owasp2025}.  Its
object-level form---broken object-level authorization (BOLA), also known as insecure direct object reference
(IDOR)---leads the OWASP API Security Top~10~\cite{owaspapi2023}, missing and incorrect authorization rank among the most
dangerous software weaknesses~\cite{cwetop25,cwe862,cwe863}, and valid IDOR reports on bug-bounty platforms keep rising
while injection and cross-site scripting decline~\cite{hackerone2025}.  A web backend serves requests on behalf of an
authenticated \emph{caller} and touches stored \emph{objects} that the request selects by an identifier.  Before an access
takes effect, some statement must establish that \emph{this} caller may perform \emph{this} operation on \emph{this}
object.  When the statement is missing, checks another object, trusts a principal that the client names, or grants
reading where the access writes, one user can read, modify, or delete another's data.

Unfortunately, a decade of detectors still leaves a gap on every front.  Unlike an injection, an authorization flaw has
no fixed shape---no source, sink, and flow between them that a rule could describe.  A check may be an owner comparison,
a query scoped by the caller's tenant, a permission predicate deep in a service, an annotation, a middleware handed to the
router three files away, or the check of a parent object that its children inherit (\cref{tab:idioms}).  \emph{Static
analyzers} therefore model one language or framework and its data layer~\cite{son2011rolecast,monshizadeh2014mace,
near2016space,huang2024bolaray,bolaz,liu2025mocguard}, and general-purpose analyzers ship IDOR queries for single
frameworks~\cite{codeqlidor} or leave the logic to rules that each project writes~\cite{semgrepidordocs}.  \emph{Dynamic
testers}~\cite{dharmaadi2025bacfuzz,rennhard2022automating,autorize} replay requests under different identities and cover
only what they manage to exercise.  \emph{LLM-based auditors}~\cite{vulnhuntr,semgrepai,claudecodesecurity,codexsecurity}
read code of any language like a human reviewer, but leave the verdict to an error-prone model, whose answers change
from run to run, flip under renaming, and flag patched code~\cite{ullah2024llms,ding2025primevul,steenhoek2024err}, and
coding agents struggle when authorization spans several files~\cite{semgrepidor}.

Worse for authorization, a model can be persuaded by code that \emph{looks} like a check.  \Cref{sec:example} details a
zero-day vulnerability that our approach detected in a popular business-intelligence platform.  The service asks a permission predicate
whether the caller may delete a personal access token---with the right library, action, and subject type---and throws
otherwise.  Shown the endpoint's code and asked for its authorization flaws, the LLM that we use reports nothing; asked
whether the deletion is restricted, it cites the predicate.  Yet the predicate restricts nothing.  The record handed to
it is assembled from the caller's own identity, so it approves every caller, and the deletion is keyed by the client's
token identifier alone.

\para{Our Approach.}  In this paper, we design and implement \sys, a neuro-symbolic framework that detects BAC in
server-side web applications without language- or framework-specific rules.  Our key insight is to separate the
\emph{proposal}, i.e., what an LLM claims about an endpoint's data accesses and the checks that guard them, from the
\emph{verdict} on whether each access is guarded, which only a symbolic verifier gives.  An LLM may flag any data
access as needing a guard, but only a symbolically verified check can clear it.  The burden of proof is thus
asymmetric.  A proposed \emph{defect} is believed, a proposed \emph{defense} counts only once proven, and a wrong
defense never clears a listed access (\cref{thm:clearance}).  LLM-assisted analyses today confirm the bugs or
specifications that the model proposes and take its dismissals as final~\cite{li2025iris,li2024llift,yang2025knighter},
the wrong way round for authorization, where what fools a model is a check that looks right (\cref{sec:example}).
Proving a defense takes what no analysis offers without per-framework models, a semantics of \emph{whose} value a check
decides about and \emph{where} it stands relative to the access.  Doing so raises three challenges.

\emph{Challenge~I---making the model's understanding checkable.}  A verdict in prose cannot be verified.  \sys asks the
model, per endpoint, for an \emph{authorization certificate}, which lists the objects that the endpoint selects from the request and
the accesses to them (\emph{obligations}, accepted as given), and the statements that restrict the caller (\emph{claims},
verified before they count), each of a fixed kind and citing a line verbatim.

\emph{Challenge~II---verifying claims without framework-specific rules.}  Whether a cited statement restricts \emph{this}
caller's access to \emph{this} object on \emph{every} path is a semantic question.  \emph{Origin tracing} determines,
for every operand of a cited statement, whose value it is---the caller's, the client's, a stored row's, or a
constant---over the syntax trees of Go, Java, JavaScript, TypeScript, and Python; a \emph{lattice-based interprocedural
dataflow analysis} then computes the caller's authority over each object on every path to the access.

\emph{Challenge~III---incompleteness on both sides.}  The model sees a slice of the program; the verifier cannot decide
every claim.  \sys keeps a three-valued verdict---cleared, a definite finding, or a lead that depends on an unverifiable
claim---and closes the loop.  That is, the verifier demands the code it found missing, states \emph{objections}, and asks whether a permission that
cleared an access is the one the operation requires.  The model answers each objection with new claims or an attack; a defense is a claim like any other and must verify.

\para{Results.}  On popular open-source applications, including Gitea, Grafana, and Strapi,
\sys detected 112 zero-day, exploitable BAC vulnerabilities, most of them in projects with over 20,000 GitHub stars.  We
have responsibly disclosed all of them; Mattermost, for example, has confirmed our report and is assigning a CVE
identifier.  On 69 pairs of vulnerable and fixed versions of an endpoint, \sys reports 85.2\% of the vulnerable versions
(macro average) where the same LLM auditing the same code reports 46.7\%, and its most confident findings alone improve recall, false alarms,
and separation at once.  Neither 2,926 fabricated claims nor 34 adversarial samples (50,903 claims)
cleared an access known to be unguarded.

\para{Contributions.}  We make the following contributions.
\begin{itemize}
  \item We design \sys, in which an LLM proposes, per endpoint, an authorization certificate listing the data accesses
  and the checks that guard them, and a symbolic verifier alone decides whether each access is guarded
  (\cref{sec:design}).
  \item We formalize the verifier---origin tracing, claim verification, a lattice-based interprocedural authority
  analysis, and a three-valued verdict---and prove that no access is cleared on the model's word, relative to five
  stated assumptions (\cref{sec:verify}).
  \item We implement \sys for five languages and evaluate it against static, BOLA, and LLM-based detectors;
  it detected 112 zero-day vulnerabilities, all disclosed (\cref{sec:eval}).  Our code is public.\footnote{\url{https://github.com/flint4bac/FLINT}}
\end{itemize}

%% file: sections/problem.tex
\section{Problem Statement}
\label{sec:problem}

\subsection{Object-Level Authorization}
\label{sec:property}

An \emph{endpoint} is a function that the framework invokes to serve an external request, e.g., a route handler, a
controller action, a view method, or an RPC or GraphQL resolver.  It acts for an authenticated \emph{caller} and performs
\emph{accesses} on stored objects.  An access $a$ has an operation $op(a) \in \{\kind{READ}, \kind{WRITE},
\kind{DELETE}, \kind{CREATE}\}$, an object $o(a)$, and \emph{key expressions} $\keys(a)$, i.e., the operands of its statement
that select which rows are touched.  An object is \emph{request-selected} when its selector is a value that the client
controls, i.e., a path, query, body, header, or cookie value, a value derived from one, or the object that the framework bound
from the URL.  Authority forms the chain $\kind{NONE} < \kind{READ} < \kind{WRITE} < \kind{FULL}$, written $0 < 1 < 2 <
3$, and an access needs $\need(\kind{READ}) = 1$ and $\need(op) = 2$ for the other operations.

\begin{definition}[Object-level authorization]
\label{def:property}
On every execution path of an endpoint, where an access to a request-selected object takes effect, the caller holds
authority of at least $\need(op)$ for that object, established earlier on the same path by a check that (1)~acts for the
authenticated caller, (2)~concerns the accessed object, (3)~ends the request when it fails, and (4)~is strong enough.
\end{definition}

\Cref{def:property} is complete mediation~\cite{saltzer1975protection} restricted to the objects that a request
selects; its violation is what CWE-639 calls an authorization bypass through a user-controlled key~\cite{cwe639}.  A
write takes effect at its statement; a read takes effect where its data \emph{leaves} the endpoint, so fetching a row,
checking its owner, and only then responding is correct.  Each clause fails in practice, e.g., with no check, a check on
the caller's own row instead of the selected one, a check that a client-chosen principal satisfies, a rejection that
does not end the request, a read permission guarding a write, an owner reference written from the request, or a batch
some of whose keys no check covers.  Function-level gaps (a route open to the wrong role) are reported apart.

\begin{table}[!htb]
\caption{How applications establish object-level authority.  Each idiom is one claim kind of \sys's certificate;
the same kind takes different syntax in every language and framework.}
\label{tab:idioms}
\centering
\footnotesize
\setlength{\tabcolsep}{4pt}
\begin{tabular}{@{}lll@{}}
\toprule
\textbf{Claim kind} & \textbf{Example idiom} & \textbf{Language} \\
\midrule
\kind{OWNER\_COMPARE} & \icode{if n.OwnerID != u.ID \{..\}} & Go \\
\kind{SCOPED\_QUERY} & \icode{objects.get(pk=pk, owner=me)} & Python \\
\kind{PERMISSION\_CALL} & \icode{ability.cannot('delete', ..)} & TS \\
 & \icode{@PreAuthorize("hasPerm..")} & Java \\
\kind{LINK} & \icode{if (c.postId !== post.id)} & JS \\
 & \icode{project.issues.get(pk=id)} & Python \\
\kind{ROLE\_GATE} & \icode{@UseGuards(AdminGuard)} & TS \\
 & \icode{r.Use(requireAdmin)} & Go \\
\bottomrule
\end{tabular}
\end{table}

\Cref{tab:idioms} lists the five ways in which applications establish authority, as \sys's certificate names them.
The same idioms appear in every language, with different syntax, receivers, and libraries, so a rule that names
\code{@PreAuthorize} or \code{ability.cannot} covers one framework at a time.  And a check often sits outside the handler,
e.g., in a service, an annotation, a middleware handed to the router, or a method that the framework invokes
(\code{canActivate}).  A detector must find these places and establish that they run before the access, on every path.

\subsection{Threat Model}
\label{sec:threat}

The attacker is any client of the application, typically an authenticated user with the lowest privilege that the
endpoint admits.  The attacker controls every value of a request and knows or guesses identifiers of other users'
objects, but cannot change code, configuration, or the database; the goal is to read, modify, delete, or create under an
object of another user, tenant, or container.  We assume that authentication is sound, i.e., the expression through which the
application reads the caller denotes the authenticated principal.  \sys analyzes source code and needs no running
instance.  Authentication flaws, injection, and capability-based sharing by design (share links) are out of scope.

%% file: sections/overview.tex
\section{Overview}
\label{sec:overview}

\subsection{A Motivating Example}
\label{sec:example}

\begin{figure}[!t]
\begin{lstlisting}[language=TypeScript]
@Middlewares([isAuthenticated, unauthorisedInDemo])
@Delete('/me/personal-access-tokens/{tokenUuid}')
async deletePersonalAccessToken(@Path() (*@\hlspan{hlred}{tokenUuid}@*): string,
                                @Request() req: Request) {
  await this.services.getPatService()
    .deletePersonalAccessToken((*@\hlspan{hlblue}{req.account}@*), tokenUuid);
  return { status: 'ok' };
}
/* PersonalAccessTokenService.ts */
async deletePersonalAccessToken(account: Account,
                                tokenUuid: string) {
  const ability = this.createAuditedAbility(account);
  if ((*@\hlspan{hlyellow}{ability.cannot('delete',}@*)           (*@\lcmt{Claim R1}@*)
      subject('PersonalAccessToken', {
        organizationUuid: (*@\hlspan{hlblue}{account.organization.uuid}@*),
        userUuid: (*@\hlspan{hlblue}{account.user.id}@*), (*@\hlspan{hlred}{tokenUuid}@*) })))
    throw new ForbiddenError('...');
  await (*@\hlspan{hlred}{this.patModel.delete(tokenUuid)}@*);   (*@\lcmt{Access A1}@*)
}
/* PersonalAccessTokenModel.ts */
async delete(tokenUuid: string) {
  await this.database('personal_access_tokens').delete()
    .where('personal_access_token_uuid', tokenUuid);
}
\end{lstlisting}
\vspace{-2pt}
\listingcaption{A motivating example: a zero-day BOLA vulnerability that \sys detected in an open-source
business-intelligence platform, simplified for explanation (identifiers shortened).  Blue marks the caller's identity, red the values selected by the
client and the access they key, and yellow the statement that the model cites as a check.}
\label{lst:example}
\end{figure}

\Cref{lst:example} presents a zero-day vulnerability that \sys detected in a popular open-source business-intelligence
platform and that we reported to the developers.  The endpoint revokes a personal access token, the credential with which
command-line tools, CI jobs, and integrations call the platform's API.

\para{Vulnerability Details.}  The controller method (lines~1--8) serves a \code{DELETE} request on a token's path
(line~2); its decorators require a session.  It passes the caller's account and the
token identifier from the path to a service (line~6).  The service builds a CASL ability for the caller (line~12) and
asks whether the caller \emph{cannot} delete a \code{PersonalAccessToken} described by a record (lines~13--16); if so, it
throws (line~17).  Otherwise, it deletes the token (line~18) through a statement keyed by the identifier alone
(lines~21--24).  The record that the predicate judges is built from the caller's organization, the caller's user
identifier, and the client's token identifier (lines~15--16).  None of its fields is read from the stored token, so
whatever rule the ability applies, it is applied to a description in which the token belongs to the caller.  (By
default, the rule that grants token management has no condition.)  Any authenticated user who
learns another user's token identifier deletes that token.  The flaw reads like a textbook check: the right
library, action, and type, and a rejection on failure.

\para{Exploitation and Impact.}  An attacker with the lowest default role sends one request with the victim's token
identifier and receives \code{200 OK} (\cref{app:exploit}).  The victim's token stops authenticating at once, which
breaks every CI job, command-line client, and automation that uses it; the deletion has no organization predicate, so the
same request reaches tokens of other tenants; and it is logged for the attacker under a generic event.

\subsection{Challenges and Our Solution}
\label{sec:solution}

Three things make this vulnerability hard to detect.  First, the check does not sit in
the handler but in a service reached through a chain of calls, and the deletion is two calls deep (lines~18 and~22).
Second, the check looks right, and only the \emph{origin} of the values handed to the predicate shows that it decides about
the caller, not the stored token.  Third, a decision must hold on the path of the deletion; a check on
another branch, or after the deletion, guards nothing.  An LLM handles the first well and the second poorly; a
static rule handles neither without knowing the framework.

\begin{figure}[!t]
  \centering
  \includegraphics[width=\columnwidth]{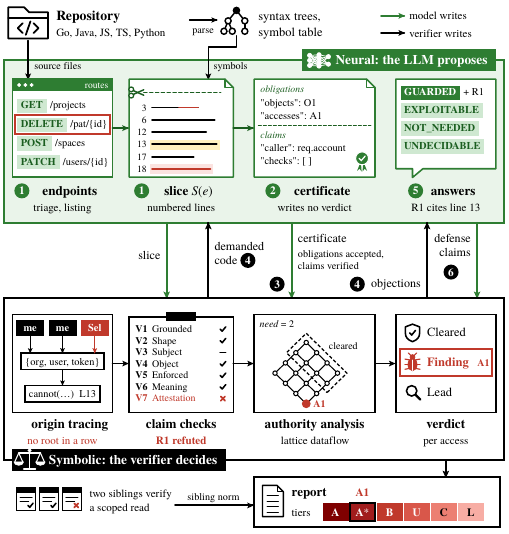}
  \caption{Workflow of \sys on \cref{lst:example}, with the steps \ding{182}--\ding{187} of \cref{sec:solution}.  The
  model (top) proposes; the verifier (bottom) decides.  Nothing the model writes clears an access unless the verifier
  confirms it in the code.}
  \label{fig:arch}
\end{figure}

\sys lets the model propose and the verifier decide (\cref{fig:arch}).  On \cref{lst:example}, it works as
follows.
\begin{itemize}[leftmargin=1.4em]
  \item[\ding{182}] \emph{Enumerate and slice.}  The model lists the controller method as an endpoint; \sys slices it
  with its decorators and the functions it calls (\cref{sec:propose}).
  \item[\ding{183}] \emph{Certify.}  The model's certificate names the caller \code{req.account}, one object (the token,
  selected by the path parameter), and one access, the deletion A1 (\cref{fig:cert}).  It cites no check.
  \item[\ding{184}] \emph{Verify.}  Origin tracing resolves the caller and the selector (\cref{sec:origin}).  With no
  claim, no authority reaches the deletion, which is a definite finding (\cref{sec:dataflow}).
  \item[\ding{185}] \emph{Demand and object.}  The verifier fetches the service and the model class that the access
  statement calls, and objects that no cited statement restricts which caller may reach the token (\cref{sec:loop}).
  \item[\ding{186}] \emph{Answer.}  The model defends the deletion with the predicate on line~13, as claim R1, which
  ``rejects the request unless the caller may `delete' a PersonalAccessToken whose [\ldots] userUuid is the caller's own
  user id''.
  \item[\ding{187}] \emph{Re-verify.}  R1 is grounded and names an authorization decision handed the caller, but every
  operand of its record has the caller's or the client's origin and none comes from the stored token, so R1 is
  refuted (\cref{sec:claims}), and the deletion stays a definite finding (\cref{fig:lattice}).
\end{itemize}
Nothing that the model said in steps \ding{183} and \ding{186} could clear the deletion; had the defense verified, the
access would have been cleared on the strength of the code it cites.

%% file: sections/design.tex
\section{Design}
\label{sec:design}

This section follows \cref{fig:arch}.  The model proposes certificates (\cref{sec:propose}), the verifier decides
every access (\cref{sec:verify}), and objections close the loop (\cref{sec:loop}).

\subsection{Proposing Authorization Certificates}
\label{sec:propose}

\para{Endpoint Enumeration.}  \sys enumerates endpoints with two model passes and resolves every answer against a
symbol table built from tree-sitter syntax trees~\cite{treesitter}.  A \emph{triage} pass shows the model short digests
of the source files and asks which may handle requests or register handlers; a \emph{listing} pass shows each selected
file as numbered lines and asks for the functions that serve requests and the statements that register them.  An
unresolvable name is dropped, so the model can name only code that exists.

\para{Demand-Driven Slicing.}  A slice is a list of sections of numbered source lines, so that every claim can cite a
location.  Unlike a program slice~\cite{weiser1984slicing}, it keeps whole functions, which the model reads better than
scattered statements~\cite{liu2024lost}, i.e., the handler with its decorators, class header, and route registration, the
definitions of the guards named there, and the functions it calls to depth two.  The slice grows on demand.  The model may
name a function whose body it needs (\code{need}), and the verifier demands code that it finds missing (\cref{sec:loop}).

\begin{figure}[!t]
\begin{promptbox}{Certificate prompt (abridged)}
You analyze ONE endpoint of a web backend and write an AUTHORIZATION CERTIFICATE: checkable facts about which stored
objects it touches and which statements restrict the caller's access to them.  \textbf{You never say whether the
endpoint is vulnerable.}  Another program verifies every fact against the code.  [\ldots]  Return only \{caller,
objects[\{selector, accesses[\{loc, quote, op, key, \ldots\}]\}], checks[\{loc, quote, kind, subject, object, form,
fail, level\}], need\}.  List every object SELECTED by client input [\ldots].  List ONLY statements that restrict which
objects THIS caller may access; authentication alone is not a check.  Every quote must be copied verbatim from the line
at its loc.  [\ldots]
\end{promptbox}
\vspace{1pt}
\begin{answerbox}{Certificate for \Cref*{lst:example} (abridged; locations are lines of the listing)}
\begin{lstlisting}[language=json,numbers=none,frame=none,basicstyle=\ttfamily\fontsize{6}{6.9}\selectfont,xleftmargin=0pt,aboveskip=0pt,belowskip=0pt]
{"caller": [{"expr": "req.account", "loc": "L6"}],
 "objects": [{"id": "O1", "type": "PersonalAccessToken",
   "selector": {"expr": "tokenUuid", "loc": "L3",
                "source": "PATH"},
   "accesses": [{"id": "A1", "op": "DELETE", "loc": "L18",
     "quote": "await this.patModel.delete(tokenUuid);",
     "key": ["tokenUuid"]}]}],
 "checks": [], "need": ["deletePersonalAccessToken"]}
\end{lstlisting}
\end{answerbox}
\caption{The certificate prompt and the model's certificate for \cref{lst:example}.  Objects and accesses are
obligations, checks are claims; there is no verdict.}
\label{fig:cert}
\end{figure}

\para{Certificates.}  For each endpoint, the model writes a certificate in a fixed JSON schema (\cref{fig:cert}), whose
two parts are treated asymmetrically.  \emph{Obligations}---the objects that the endpoint selects from client input and
the accesses to them, each with a location, a verbatim quote, an operation, and key expressions---are accepted as given,
since a spurious obligation can only add a finding.  \emph{Claims}---the caller expression and the checks---count only once
verified.  A check claim cites a statement and gives its kind (\cref{tab:idioms}), subject and object expressions,
\emph{form} (\kind{REJECT} if failure ends the request, \kind{POSITIVE} if the access sits in the passing branch,
\kind{THROWS} if the call aborts, \kind{FILTER} if the query excludes other callers' rows), failing statement, and level.  A
coverage list that the schema asks for (which check covers which access) is never read, because the verifier computes
coverage itself.  For \cref{lst:example}, the model lists the deletion, cites no check, and asks for the service's body.

\subsection{Verifying Certificates}
\label{sec:verify}

The verifier decides, for every access of a certificate, whether the code establishes \cref{def:property}; it never
asks the model anything.  We derive each of its steps on \cref{lst:example}.

\subsubsection{Language and Notation}
\label{sec:notation}

We present the verifier over a core language that captures the constructs it interprets in all five
languages.  A program is an endpoint's slice, i.e., an entry function $E$ and the functions it may call, over
\[
\begin{array}{r@{\;}c@{\;}l}
e & ::= & x \mid c \mid e.f \mid e_0.m(e_1,\ldots,e_n) \mid e_1 \odot e_2 \\
  &     & \mid \{f_1{:}\,e_1,\ldots,f_n{:}\,e_n\} \\
s & ::= & x := e \mid e.f := e \mid e \mid s_1;s_2 \\
  &     & \mid \textbf{if}\ e\ \textbf{then}\ s_1\ \textbf{else}\ s_2 \mid \textbf{while}\ e\ \textbf{do}\ s \\
  &     & \mid \textbf{return}\ e \mid \textbf{throw}\ e
\end{array}
\]
with variables $x$, constants $c$, fields $f$, and method or function names $m$.  Per-language tables of syntax-tree node
types map each language onto this core (\cref{sec:impl}).  For example, destructuring becomes field reads, an out-parameter an
assignment, \code{try}/\code{catch} a branch, and a decorator, annotation, or route middleware a guard statement at the
entry of $E$.  \Cref{tab:notation} summarizes our notation.

\begin{table}[!htb]
\caption{Notation used in \sys's verifier.}
\label{tab:notation}
\centering
\footnotesize
\setlength{\tabcolsep}{4pt}
\renewcommand{\arraystretch}{1.04}
\begin{tabular}{@{}p{0.25\columnwidth}p{0.69\columnwidth}@{}}
\toprule
\textbf{Origins} & \textbf{Description} \\
\midrule
$r = (t, b, \pi)$ & Root of kind $t \in \{\rootP, \rootS, \rootG, \rootL, \rootU\}$ (parameter, receiver, global, literal
or unset field, unknown), base $b$, path $\pi$ \\
$\pi$ & Fields $f$, accessors $m(\bar c)$, derivations $\dstep{m}$, keyed lookups $\kstep{m}$ \\
$\orig(e)$ & The roots of expression $e$ (\cref{fig:rules}) \\
$\Rc$,\ $\ext{R}$ & Caller roots; roots that extend a root of $R$ \\
$\Client(r)$ & $r$ is client-controlled request data \\
$\Sel(o)$, $\Row(o)$ & Roots of $o$'s selector; of its fetched rows \\
$r \lin o$ & $r$ lies on the lineage of object $o$ \\
\midrule
\textbf{Certificate} & \\
\midrule
$O$, $A$, $K$ & Objects, accesses (obligations); check claims \\
$op(a)$, $o(a)$, $\keys(a)$ & Operation, object, key expressions of access $a$ \\
$s_k$, $\st(k)$ & Statement cited by $k$; status $\mValid$, $\mRefuted$, or $\mUnverified$ \\
$\Obj(k)$, $\lvl(k)$ & Objects that check $k$ concerns; level it grants \\
\midrule
\textbf{Authority} & \\
\midrule
$\Lambda = \{0,1,2,3\}$ & \kind{NONE} $<$ \kind{READ} $<$ \kind{WRITE} $<$ \kind{FULL} \\
$\sigma = (\alpha, \Links)$ & State, where $\alpha$ maps $O \cup \{\#k, @gate, @admin\}$ to $\Lambda$ and $\Links$ holds link facts \\
$\sqcup$, $\sqcap$, $\bot$, $\top$ & Join, meet, least and greatest state \\
$\sem{s}\sigma$;\ $\Tc(c)$, $\Fc(c)$ & State after $s$; facts on the true/false side of $c$ \\
$\contrib(k)$ & Facts contributed by a \mValid{} check $k$ \\
$\auth(\sigma)(o)$, $\need(op)$ & Authority held for $o$; authority that $op$ needs \\
$\sigma^-_a$, $\sigma^+_a$ & Pessimistic, optimistic state where $a$ takes effect \\
\bottomrule
\end{tabular}
\end{table}

\subsubsection{Origin Tracing}
\label{sec:origin}

\begin{figure*}[t]
\centering
\scriptsize
\newcommand{\rul}[3]{$\inferrule*[right=\rn{#1}]{#2}{#3}$}%
\begin{tabular}{@{}l@{\hspace{4.5em}}l@{}}
\rul{Param}{x \text{ is a parameter of } E}{x \Downarrow \{(\rootP, x, \epsilon)\}} &
\rul{Bind}{x \text{ is parameter } i \text{ of } g \neq E \\ e_{j,i} \Downarrow R_j \text{ for each call }
  e_{j,0}.g(e_{j,1},\ldots) \text{ of } g}{x \Downarrow \textstyle\bigcup_j R_j} \\[5pt]
\rul{Def}{x := e_j \text{ in the function of } x \\ e_j \Downarrow R_j}{x \Downarrow \textstyle\bigcup_j R_j} &
\rul{Base}{ }{\mathtt{this} \Downarrow \{(\rootS, \mathtt{this}, \epsilon)\} \quad g \Downarrow \{(\rootG, g, \epsilon)\}
  \quad c \Downarrow \{(\rootL, c, \epsilon)\}} \\[5pt]
\rul{Field}{e \Downarrow R \\ e \text{ is bound to no record literal}}{e.f \Downarrow R \cdot f} &
\rul{Rec}{e \equiv \{\ldots, f{:}\,e_f, \ldots\} \\ e_f \Downarrow R}{e.f \Downarrow R} \\[5pt]
\rul{Unset}{e \equiv \{f_1{:}\,e_1, \ldots, f_n{:}\,e_n\} \\ f \notin \{f_1,\ldots,f_n\}}{e.f \Downarrow \{(\rootL, \bot,
  f)\}} &
\rul{Derive}{e_0 \Downarrow R_0 \\ e_i \Downarrow R_i \ (1 \le i \le n) \\ \textstyle \bigcup_{i} R_i \not\subseteq
  \{(\rootL,\cdot,\cdot)\}}{e_0.m(e_1,\ldots,e_n) \Downarrow \textstyle\bigcup_i R_i \cdot \dstep{m} \;\cup\; R_0 \cdot
  \kstep{m}} \\[5pt]
\rul{Acc}{e_0 \Downarrow R_0 \\ c_1,\ldots,c_n \text{ constants}}{e_0.m(c_1,\ldots,c_n) \Downarrow R_0 \cdot m(\bar c)} &
\rul{Ret}{e_0.m(\bar e) \text{ calls } g \text{ of the slice} \\ \textbf{return}\ e'_j \text{ in } g \\ e'_j \Downarrow
  R'_j}{e_0.m(\bar e) \Downarrow \textstyle\bigcup_j R'_j} \\[5pt]
\rul{Merge}{x.m(e) \text{ with } m \text{ a container mutator} \\ e \Downarrow R}{x \Downarrow R \cdot \dstep{merge}} & \\
\end{tabular}
\vspace{-2pt}
\caption{Origin-tracing rules.  $e \Downarrow R$: expression $e$ has the roots $R$, and $\orig(e)$ is the union over all
derivations; $e \equiv \rho$: $e$ is bound to the record literal $\rho$ through definitions and call sites; $R \cdot
\delta$ appends step $\delta$ to the path of every root in $R$.  A call whose inputs carry no data root has the unknown
root $(\rootU, m, \epsilon)$.}
\label{fig:rules}
\end{figure*}

Verification asks, for every operand of a cited statement, \emph{whose value is this}.  Taint
analysis~\cite{livshits2005finding,tripp2009taj,arzt2014flowdroid} asks whether a value \emph{may} come from a source;
here the kind of every origin matters, because a check is sound only if one operand is the caller and another the
object.  \sys computes the \emph{origin} of an expression as a set of roots by the rules of \cref{fig:rules}.

\para{Roots and Rules.}  A root $(t, b, \pi)$ says that a value is reached from base $b$ of kind $t$ along path $\pi$, i.e., a
sequence of steps, each a field $f$; an \emph{accessor} $m(\bar c)$, a call with constant arguments such as \code{ctx.Param("id")} (\rn{Acc}); a
\emph{derivation} $\dstep{m}$, ``passed through call $m$''; or a \emph{keyed lookup} $\kstep{m}$ on the receiver of a
call that is given data---\code{repo.find(id)} yields \emph{some} element of \code{repo}, selected by \code{id}
(\rn{Derive}).  Identifiers take the union of their definitions (\rn{Def}); a callee's parameter takes the origins of its
arguments at every call site that the slicer resolved or whose arity and receiver fit (\rn{Bind}); and a call of a slice
function also takes the origins of its return expressions (\rn{Ret}).  Field reads are field-sensitive through record
literals (\rn{Rec}), and a field that no literal sets is \emph{unset} (\rn{Unset}).  Two further rules keep identities
apart.  First, \code{this} in a method called on another object denotes that object, never the caller.  Second, the request
parameter of a guard that the framework invokes denotes the handler's request.

\para{Caller, Client, and Objects.}  We define four notions.
\begin{itemize}
  \item The \emph{caller roots} $\Rc = \orig(x_{\mathit{me}})$ are the origins of the certificate's caller expression
  (rejected if it lies on a request selector); a root in $\ext{\Rc}$ is \emph{caller-derived}.
  \item $\Client(r)$ holds when $r$ is not caller-derived and lies on a declared selector, is read through the accessor
  that such a selector uses, or is read through a generic request accessor (\code{body}, \code{query}, \code{params}).
  \item An object $o$ is \emph{denoted} by $\Sel(o)$, its selector's roots, and $\Row(o)$, the roots of the variables
  holding its rows.
  \item $r \lin o$ ($r$ lies on the \emph{lineage} of $o$) if, for some $d \in \Sel(o) \cup \Row(o)$, $r = d$; $r$
  extends $d \in \Row(o)$ (a row's field); $r = d \cdot \dstep{m} \cdots$ for $d \in \Sel(o)$ (the selector
  passed on); or $d = r \cdot \dstep{m} \cdots$ (what $d$ was derived from).
\end{itemize}
Keyed steps never establish identity, because two rows fetched from one table are different objects, and a container
from which the caller or several selectors are read (the request itself) concerns none of them.

\begin{figure}[!t]
  \centering
  \includegraphics[width=\columnwidth]{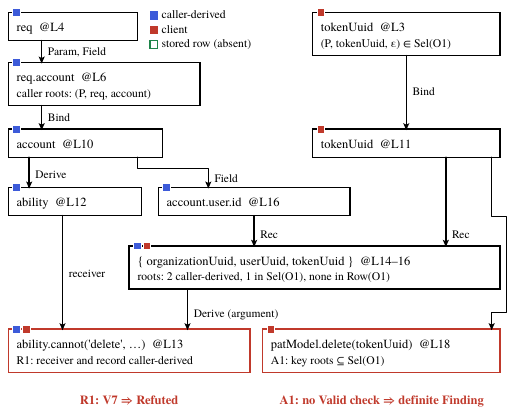}
  \caption{Origin graph of \cref{lst:example}: nodes are expressions with their line, edges carry the deriving rule of
  \cref{fig:rules}, and flags show root kinds.  The predicate's arguments carry caller and client roots but none of a
  stored row, which refutes claim R1 (V7 in \cref{tab:spec}).}
  \label{fig:origin}
\end{figure}

\begin{example}[Origins in \cref{lst:example}]
\label{ex:origin}
The caller is \code{req.account} (line~6), so $\Rc = \{(\rootP, \mathtt{req}, \mathtt{account})\}$.  The service's
parameters are bound on line~6 (\rn{Bind}), so the fields on lines~15--16 have the caller-derived roots
$(\rootP, \mathtt{req}, \mathtt{account}{\cdot}\mathtt{user}{\cdot}\mathtt{id})$ and $(\rootP, \mathtt{req},
\mathtt{account}{\cdot}\mathtt{organization}{\cdot}\mathtt{uuid})$, and the client root $(\rootP, \mathtt{tokenUuid},
\epsilon) \in \Sel(O_1)$.  The receiver \code{ability} (line~12) is caller-derived by \rn{Derive}.  No argument of the
call on line~13 has a root in $\Row(O_1)$, i.e., nothing given to the predicate was read from the stored token
(\cref{fig:origin}).
\end{example}

\subsubsection{Claim Verification}
\label{sec:claims}

\begin{table*}[t]
\caption{Specification of claim verification.  A claim $k$ cites statement $s = s_k$ with operands $\ops(s)$.  A
violated condition makes the claim \REFUTED (the code contradicts it) or \UNVERIFIED (the code neither confirms nor
contradicts it); only a \VALID claim contributes authority.}
\label{tab:spec}
\centering
\footnotesize
\setlength{\tabcolsep}{4pt}
\renewcommand{\arraystretch}{1.1}
\begin{tabular}{@{}llp{0.535\textwidth}l@{}}
\toprule
\textbf{Rule} & \textbf{Kinds} & \textbf{Condition on the cited statement $s$} & \textbf{If violated} \\
\midrule
V1 Grounded & All & The quote of $k$ occurs at its cited line ($\pm 2$ lines, whitespace-normalized) & \mUnverified \\
V2 Shape & All & $s$ is a comparison or query (\kind{OWNER\_COMPARE}, \kind{SCOPED\_QUERY}) or a call or annotation
(\kind{PERMISSION\_CALL}), with $\mathit{form}(k) \neq \kind{NONE}$; kind and form are re-read from $s$ & \mRefuted\ (\why{R-SHAPE}) \\
V3 Subject & \kind{OWNER\_COMPARE}, & For every alternative $\alpha$ of $s$ that the client can select, $\exists x \in
\ops_\alpha(s)$ in a selecting position with $\orig(x) \cap \ext{\Rc} \neq \emptyset$, & \mRefuted\ (\why{R-SUBJ}) \\
 & \kind{SCOPED\_QUERY} & and no subject operand has a root $r$ with $\Client(r)$ & \\
V4 Object & All & $\Obj(k) = \{\, o \in O \mid \exists x \in \ops(s),\ r \in \orig(x).\ r \lin o \,\}$; if
$\Obj(k) = \emptyset$, $k$ is a gate ($@gate$, or $@admin$ if $s$ names an admin role) & Gate only \\
V5 Enforced & By form & For \kind{REJECT}, the failing branch of $s$ ends the request (\textbf{return}, \textbf{throw}, or
a no-return call) or, in a loop, the iteration (\textbf{continue}, \textbf{break}); for \kind{THROWS}, the callee has an aborting exit; for \kind{FILTER}, the access runs the query value
that $s$ builds; for \kind{POSITIVE}, the access lies in the passing branch & \mRefuted\ (\why{R-NOTHROW}) \\
V6 Meaning & \kind{PERMISSION\_CALL} & ($s$ is tested by a branch $\wedge\ \exists x \in \ops(s).\ \orig(x) \cap
\ext{\Rc} \neq \emptyset$) $\vee$ the callee or annotation names an authorization concept & \mUnverified\
(\why{R-OPAQUE}) \\
V7 Attestation & \kind{PERMISSION\_CALL} & $\neg\big(\exists$ record literal $\rho$ among the arguments of $s$ with
$\orig(\rho) \cap \ext{\Rc} \neq \emptyset \wedge \orig(\mathit{recv}(s)) \cap \ext{\Rc} \neq \emptyset \wedge$ no
argument root reads a field of a fetched row$\big)$ & \mRefuted\ (\why{R-ATTEST}) \\
\midrule
Level & All & $\lvl(k) = 3$ for \kind{OWNER\_COMPARE}; the declared level for \kind{PERMISSION\_CALL}, capped at 1 if its
name or constant arguments say only \emph{view} or \emph{read}; for \kind{SCOPED\_QUERY}, 3 if it is the only guard of its
object and the declared level otherwise & --- \\
\bottomrule
\end{tabular}
\end{table*}

A check claim becomes \VALID only if the code satisfies every condition of \cref{tab:spec}.  The conditions mirror the
four clauses of \cref{def:property}.  V3 and V6 establish that the check acts for the caller (clause~1), V4 that it
concerns an object (clause~2), V5 that it ends the request or restricts the query (clause~3), and the level rule that it
is strong enough (clause~4); V1 and V2 tie the claim to real code of the claimed shape.  Transparent kinds (owner
comparisons, scoped queries, links) are decided by their operands.  An opaque predicate's \emph{meaning} rests on what
the code says about it (V6), and its level on its declaration, lowered but never raised by its name.  A \kind{LINK}
carries a parent's authority to its child (\cref{sec:dataflow}).

\para{Hardening.}  Each condition is easy to satisfy with a misleading claim.  Every way in which a model made a false
claim pass---in the adversarial rounds of \cref{sec:rq4}, under fault injection, or in probing by hand---became a
refinement of V3--V7 with a pair of conformance tests (the safe pattern and its nearest unsafe variant), following four
principles.  \emph{Position}---an operand counts in the role that its syntax gives it; the caller scopes a query only in a
selecting position, not as a written value or under a negation.  \emph{Use}---a call is a guard only if its result is used
as claimed.  \emph{Identity}---the caller is what is read through the caller's accessor, not a value computed from the
caller and client data together, and not a record assembled from the caller's identity in place of the stored object
(V7).  A predicate of the repository that is handed a request value as its principal, and reads no principal itself,
decides nothing about the caller.  A path joined from the caller's base directory and a request segment is refuted unless
the code tests it for confinement, since a segment such as \code{..} leaves the base.  \emph{Evidence over word}---where the code shows something, the claim's word yields.

\subsubsection{Authority Analysis}
\label{sec:dataflow}

Whether a \VALID check holds where an access takes effect is decided by a forward, must-style dataflow
analysis~\cite{kildall1973unified,kam1977monotone} over the structured syntax trees of the slice, whose domain of
authority levels is in the tradition of lattice models of secure information flow~\cite{denning1976lattice}.

\begin{definition}[Authority lattice]
\label{def:lattice}
Let $\Theta = O \cup \{\#k \mid k \in K\} \cup \{@gate, @admin\}$, where $\#k$ stands for what a \VALID check concerns
when that is not a listed object.  A state $\sigma = (\alpha, \Links)$ consists of $\alpha : \Theta \to \Lambda$ and a set
$\Links$ of link facts, ordered by $\sigma_1 \sqsubseteq \sigma_2$ iff $\alpha_1 \le \alpha_2$ pointwise and $\Links_1
\subseteq \Links_2$.  Join $\sqcup$ and meet $\sqcap$ are the pointwise maximum and minimum (union and intersection on
$\Links$); $\bot$ maps everything to 0 with no links, and $\top$ everything to 3 with all links.
\end{definition}

As a product of chains and a powerset, $(\Sigma, \sqsubseteq)$ is a complete lattice of finite height.  A \VALID check
contributes $\contrib(k) = (\{o \mapsto \lvl(k) \mid o \in \Obj(k)\}, \emptyset)$; a gate contributes to $@gate$ or
$@admin$, and a link adds itself to $\Links$.  \Cref{tab:transfer} gives the transfer functions.  The entry state of $E$
joins the contributions of its \VALID declarative guards (decorators, annotations, class headers, and functions
registered before the route), so that a guard outside the handler holds on all of its paths.

\begin{table}[!htb]
\caption{Transfer functions of the authority analysis.  $\top$ is the state of a path that has ended; it is the
identity of $\sqcap$, so ended paths constrain nothing.}
\label{tab:transfer}
\centering
\footnotesize
\setlength{\tabcolsep}{3pt}
\renewcommand{\arraystretch}{1.12}
\begin{tabular}{@{}p{0.40\columnwidth}p{0.555\columnwidth}@{}}
\toprule
\textbf{Statement $s$} & $\sem{s}\sigma$ \\
\midrule
$s_1; s_2$ & $\sem{s_2}(\sem{s_1}\sigma)$ \\
$s$ cites \VALID $k$, form \kind{THROWS} or \kind{FILTER} & $\sigma \sqcup \contrib(k)$, after $s$ completes \\
$\textbf{if}\ c\ \textbf{then}\ s_1\ \textbf{else}\ s_2$ & $\sem{s_1}(\sigma \sqcup \Tc(c)) \sqcap \sem{s_2}(\sigma
\sqcup \Fc(c))$ \\
$\textbf{return}\ e$, $\textbf{throw}\ e$ & $\top$ (the path ends) \\
$\textbf{while}\ c\ \textbf{do}\ s_1$ & $\sigma$; the body is analyzed from $\sigma$ \\
Call of slice function $g$ & $\bigsqcap_j \sigma_j$ over the returns $j$ of $g$ compatible with the caller's test of
the result; $g$ inlined to depth 3 \\
Any other $s$ & $\sigma$ \\
\bottomrule
\end{tabular}
\end{table}

\para{Conditions.}  A branch condition is put in negation normal form over \emph{atoms}, and $\Tc(c_1 \wedge c_2) =
\Tc(c_1) \sqcup \Tc(c_2)$, $\Tc(c_1 \vee c_2) = \Tc(c_1) \sqcap \Tc(c_2)$, with $\Fc$ dual.  A \VALID check atom
contributes $\contrib(k)$ on its passing side---decided by the operator of a comparison, and for an opaque predicate by
the side without the cited rejection---and $\bot$ on the other.  An atom over client data (\code{req.query.force})
contributes $\bot$; an atom over caller-derived data (\code{me.isAdmin}) or a stored attribute of an accessed object
(\code{doc.public}) contributes $\neutral$, the identity of both $\sqcup$ and $\sqcap$.  Hence \code{owner == me ||
req.query.force} yields $\contrib(k) \sqcap \bot = \bot$.  In other words, an alternative that the client controls destroys the
fact, while one reserved to privileged callers leaves it intact.

\para{Calls and Loops.}  Calls into the slice are inlined, a bounded form of interprocedural
analysis~\cite{sharir1981two,reps1995ifds}.  A callee is summarized by its returns, each with its state and a
\emph{status vector} recording which returned values are certainly truthy or falsy, so that a caller that tests
\code{err != nil} continues with the compatible returns only; a function that returns a check's result contributes the
check on its truthy return only.  A loop body runs from the state before the loop, which is also the state after it.
Since every transfer function only adds facts and merges take meets, no fixpoint iteration is needed.

\para{Effects.}  A write, deletion, or creation takes effect at its statement; a read where its data is
\emph{released}---returned, handed to a call that can reach the client, or stored into non-local state, by a
call-site-sensitive data dependence from the access's result.  The authority for $o$ at an effect is
\begin{multline*}
\auth(\sigma)(o) = \max\big(\alpha(o),\ \max\nolimits_{(o,p) \in \Links} \auth(\sigma)(p),\\
\max\nolimits_{p \in \mathit{src}(a)} \alpha(p)\big),
\end{multline*}
where $\mathit{src}(a)$ are the objects from whose stored data the access's key is derived.  Key coverage is a
\emph{must} property, $\KeyCovered(a, \sigma)$: every client root of every key must lie on the lineage of an operand of
an applied \VALID check or derive from stored data of an authorized object; roots that enter one lookup together are
joint (\code{find(id, tenant)}), and roots of alternative sources each need cover (\why{R-KEY}).  A written owner
reference taken from client data is a finding (\why{R-REF}).

\begin{figure*}[t]
  \centering
  \includegraphics[width=\textwidth]{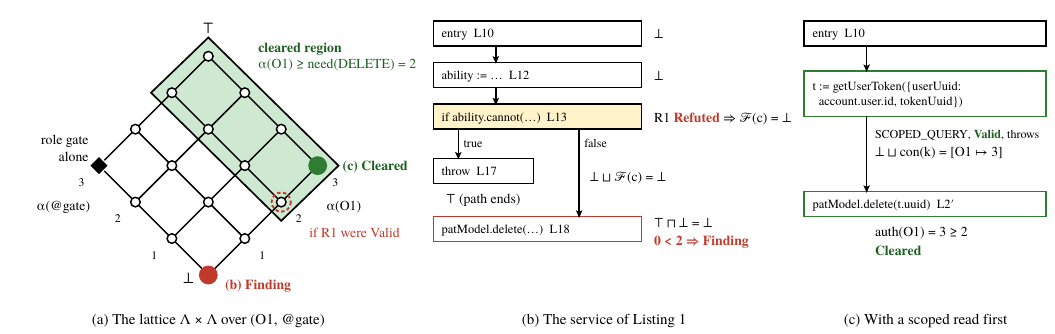}
  \caption{Authority analysis of \cref{lst:example}.  (a)~The lattice $\Lambda \times \Lambda$ over $(O_1, @gate)$;
  the shaded up-set holds the states that clear the deletion, $\alpha(O_1) \ge \need(\kind{DELETE}) = 2$, and a role gate
  alone stays outside it.  (b)~The states along the service, where refuted R1 contributes nothing, so the deletion is reached in
  $\bot$ under both evaluations, a definite finding.  (c)~The states along a version that first reads the token through a
  query scoped by the caller.}
  \label{fig:lattice}
\end{figure*}

\begin{example}[Authority for the deletion in \cref{lst:example}]
\label{ex:flow}
The decorators on line~1 authenticate but name no object or role, so the entry state is $\bot$.  The \code{if} on
line~13 has one atom, the call cited by R1; its true side throws (line~17) and yields $\top$, and since R1 is \REFUTED,
its false side is $\bot \sqcup \Fc(c) = \bot$.  The merge gives $\top \sqcap \bot = \bot$, so $\auth(\sigma)(O_1) = 0 <
\need(\kind{DELETE}) = 2$ under both evaluations (\cref{fig:lattice}).  Had the service first read the token through a
query scoped by \code{account.user.id}, as its sibling that rotates tokens does, that \kind{SCOPED\_QUERY} would be
\VALID and the only guard of $O_1$, and the state after it would be $\{O_1 \mapsto 3\}$---the deletion would be cleared.
\end{example}

\subsubsection{Verdict and Guarantees}
\label{sec:verdict}

Not every claim is decidable.  A predicate defined in a library may name no authorization concept, and a callee may be
missing from the slice.  \sys therefore evaluates the analysis twice.

\begin{definition}[Three-valued verdict]
\label{def:verdict}
Let $\sigma^-_a$ be the state where access $a$ takes effect when only \VALID claims contribute (pessimistic), and
$\sigma^+_a$ the state when, in addition, every \UNVERIFIED claim not made in defense is assumed as stated and \VALID
checks of unknown position are placed first (optimistic).  Access $a$ on $o = o(a)$ is
\begin{itemize}
  \item \Cleared if $\auth(\sigma^-_a)(o) \ge \need(op(a)) \wedge \KeyCovered(a, \sigma^-_a) \wedge \neg\Broken(a)$;
  \item a definite \Finding if $\auth(\sigma^+_a)(o) < \need(op(a)) \vee \Broken(a)$;
  \item a \Lead otherwise, i.e., it depends on a claim that the code cannot decide.
\end{itemize}
\end{definition}

$\Broken(a)$ keeps a write uncleared when the code itself makes an authorization decision about the object that is
refuted for its subject, unless a \VALID decision of the needed level about the caller covers it.

\begin{lemma}[Claims that do not verify cannot clear]
\label{lem:inert}
Adding claims that are not \VALID to a certificate never clears an access.  At most, an \UNVERIFIED claim turns a
definite finding into a lead; a defense claim cannot even do that.
\end{lemma}

\noindent\emph{Proof sketch.}  \REFUTED claims enter neither evaluation and serve only $\Broken$, which only keeps an
access uncleared; \UNVERIFIED claims enter only $\sigma^+_a$; defense claims enter neither unless \VALID.  The transfer
functions are monotone, so $\sigma^-_a$, and with it the cleared set, is unchanged.\hfill$\square$

\begin{theorem}[Clearance is decided by code]
\label{thm:clearance}
If \sys clears access $a$, then on every path of the analyzed program that reaches the effect of $a$, some statement $s$
precedes it such that, as established from the code alone, (i)~$s$ ends the request when it fails or filters the query
that $a$ runs; (ii)~the caller is an operand of $s$ on every client-selectable alternative, or $s$ is a predicate
accepted under V6 and V7; (iii)~an operand of $s$ lies on the lineage of every client source of $\keys(a)$, or the key
derives from stored data of an object so covered; and (iv)~$\lvl(s) \ge \need(op(a))$.
\end{theorem}

\noindent\emph{Proof sketch.}  In the pessimistic evaluation, authority enters a state only by a join with
$\contrib(k)$ of a \VALID claim $k$ where $k$ passes (V5), and a meet keeps a fact only if it holds on every incoming path
that has not ended.  Hence $\auth(\sigma^-_a)(o) \ge \need(op(a))$ implies that every path to the effect passes a \VALID
check of sufficient level concerning $o$, its parent, or the source of its key; V1--V7 establish (i), (ii), and (iv),
and $\KeyCovered$ establishes (iii).\hfill$\square$

The certificate contributes only the location and kind of $s$, and for an opaque predicate or a scope its level.  The
theorem is relative to five assumptions, namely (A1)~a call that the analysis inlines is a call of that function; (A2)~the
declared caller expression, once shown not to be client data, denotes the authenticated principal; (A3)~a predicate
accepted under V6 and V7 decides whether the caller may access its object operands, at its declared level unless its
name says less; (A4)~a verified scope or link grants the level that the certificate states, unless the code contradicts
it; and (A5)~guards named by the endpoint's declaration or registration run before the handler.  Outside the theorem are
accesses that the model did not list, reflection, flows through the store, and authorization by possession of a
capability (share links, signed URLs).

\para{Termination.}  The dataflow analysis visits each statement of each inlined clone once.  Origin tracing is
exponential in the worst case; an endpoint that exceeds 60\,s is evaluated again with a budget of 3,000 evaluations,
beyond which origins are unknown, and such a bounded result never clears.

\subsection{Closing the Loop}
\label{sec:loop}

A definite finding may be a guard that the model missed, and a lead may hide a real flaw, so the verifier answers back.

\para{Demand and Objections.}  For every endpoint with an uncleared access, the verifier adds the code that it found
missing, resolved from the code and never from a name that the model wrote.  That is, it adds the function that an uncleared access
statement calls (resolved by arity and receiver), the body of an undecided predicate, the definitions
behind the decorators, and the \emph{registration scope}---statements that hand functions to the router before the route
(\code{router.use(mw)}).  It then states one \emph{objection} per uncleared access, giving its location and the reason in terms
of the code (no restriction cited, only gates, an uncovered key, not on every path, a weaker level, a subject that is not
the caller, a self-attested record), never whether the endpoint is vulnerable.  The model answers each objection with
\kind{GUARDED} and new claims, \kind{EXPLOITABLE} with an attack, \kind{NOT\_NEEDED} with a reason, or \kind{UNDECIDABLE}
with the code it was not shown.  \kind{GUARDED} claims are verified like any other and count only if \VALID
(\cref{lem:inert}).

\para{Challenging the Level.}  Verification takes the level of a permission predicate from its declaration and name
(A3), so a name that says \emph{read} or \emph{view} clears a read, although applications define finer ones,
e.g., to see a workspace, logs, or secrets.  For every access cleared by such predicates alone, \sys asks the model in
three independent samples which permission the operation requires and whether the checked one implies it; if the
majority names a finer one and an attack, the access stays uncleared, a finding conceded with an attack.
Like an obligation, the challenge only adds findings.

\para{The Application's Own Norm.}  An endpoint that checks \emph{something}, but less than the application checks
elsewhere for the same type of object, passes the steps above.  \sys inventories, per object type, the guards
verified on the application's endpoints, under a signature read from the code (a predicate with its constant arguments,
or the compared owner field).  An endpoint deviates when at least two siblings verify a strictly stronger guard on a type
that it accesses and its own code neither verifies, claims, nor mentions it; the model must then cite the access that the
guard should protect, and can only add a finding.  For \cref{lst:example}, two siblings that list and rotate tokens
verifiably fetch them with caller-scoped queries, and the deletion does not.

\para{Report.}  An endpoint is reported if an access stays uncleared, in the best tier of its definite findings, i.e.,
\tier{A}, conceded by the model with an attack; \tier{A*}, below a verified norm of its siblings; \tier{B}, defended,
with the defense refuted; \tier{U}, undecided for lack of code; \tier{C}, waived as not needed; and \tier{L}, a lead.  The
tiers order the report; none removes an entry.  \Cref{lst:example} is reported in tier~A*.

%% file: sections/implementation.tex
\section{Implementation}
\label{sec:impl}

We implemented \sys in 14.1\,K lines of Python, namely the verifier (8.6\,K), the neural pipeline (3.2\,K), and
drivers for the objection round, the need challenge, the sibling norm, and fault injection (2.3\,K).  The only language-specific code is a table of tree-sitter~\cite{treesitter} node types per language
family (Python; JavaScript, TypeScript, and TSX; Go; Java), which maps each language onto the core language of
\cref{sec:notation}.  No rule names an application, a framework, a library, an ORM, or a permission API.  A few vocabularies are applied
to code text only: request accessors, calls that do not return, container mutators, and words that name an
authorization concept (only as evidence for V6).  Test files are not analyzed.  Each verification
rule is a switch with a pair of conformance tests; the 283 tests (Go~95, Python~56, JavaScript~54, TypeScript~53, Java~25)
run in 11\,s.

\sys uses DeepSeek-V4.1-Flash, a successor of DeepSeek-V3~\cite{deepseekv3}, through its
API~\cite{deepseekapi} with thinking at the \emph{low} effort setting and JSON output.  Every model answer is cached under
a hash of the prompt version, model, effort, and messages, so that every result in \cref{sec:eval} can be reproduced
from the cache without the model.

%% file: sections/evaluation.tex
\section{Evaluation}
\label{sec:eval}

We evaluate \sys on five research questions (RQs).
\begin{itemize}
  \item \RQ{1}{Zero-day}---How many zero-day BAC vulnerabilities does \sys detect in popular applications?
  \item \RQ{2}{Comparison}---How does \sys compare with static, BOLA, and LLM-based detectors on known vulnerabilities and
  their fixes?
  \item \RQ{3}{Ablation}---How much do authority analysis, claim verification, and the loop contribute?
  \item \RQ{4}{Soundness of Clearance}---Can a wrong or an adversarial model make \sys clear an unguarded access?
  \item \RQ{5}{Practicality}---What does \sys report on whole applications, and at what time and cost?
\end{itemize}

\subsection{Experimental Setup}
\label{sec:setup}

\begin{table}[!htb]
\caption{Our datasets and the research questions that use them.}
\label{tab:datasets}
\centering
\footnotesize
\setlength{\tabcolsep}{3pt}
\begin{tabular}{@{}lp{0.765\columnwidth}l@{}}
\toprule
\textbf{Set} & \textbf{Content; languages} & \textbf{RQ} \\
\midrule
\textsc{Zero} & 57 popular open-source applications, the subjects of our search for zero-days; JS/TS~21, Py~14, Go~13, Java~9 & 1 \\
\textsc{Pairs} & 69 pairs, each an endpoint before and after a public BAC fix, from 35 repositories; Go~26, Java~22, JS/TS~13, Py~8 & 2--4 \\
\textsc{Hunt} & 50 vulnerabilities that we found and reproduced (43 object-level, 41 with a function at the flaw); JS/TS~20, Py~11, Java~6, Go~4 & 2--4 \\
\textsc{Neigh} & The 41 \textsc{Hunt} endpoints and their 227 siblings in the same files, in 26 applications & 2, 3, 5 \\
\textsc{Repo} & 5 whole applications: 831 functions, 786 handlers; Go~3, JS/TS~2 & 5 \\
\textsc{Near} & 33 fixes of \textsc{Pairs}, each mutated so that its attack works again; Java~20, Py~8, JS/TS~3, Go~2 & 2 \\
\bottomrule
\end{tabular}
\end{table}

\para{Datasets.}  We built the six sets of \cref{tab:datasets} ourselves, from public fixes, from the
vulnerabilities we found, and from the applications we audited.  Recent BAC detectors were each evaluated on PHP or Java alone~\cite{huang2024bolaray,dharmaadi2025bacfuzz,liu2025mocguard}; ours span the five languages of the popular multi-user applications we target and pair known vulnerabilities with their fixes, to count false alarms on repaired code.  \textsc{Zero} consists of popular open-source web
applications with multi-user permission models (a median of 23.6K GitHub stars), at the commits that we audited in October and November 2026.  \textsc{Pairs} holds code that its
developers deemed vulnerable and then repaired, and \textsc{Hunt} vulnerabilities that were unknown when we found them,
each reproduced with two accounts; \textsc{Neigh}, \textsc{Repo}, and \textsc{Near} ask what a detector says next to a
vulnerability, on whole applications, and on near misses.  As is common in vulnerability discovery, we built \sys while
running it on these sets, turning each crash, parse error, or missed flow into a general rule.

\begin{table}[!htb]
\caption{Baselines and the dataset languages they support.}
\label{tab:baselines}
\centering
\footnotesize
\setlength{\tabcolsep}{2.4pt}
\begin{tabular}{@{}llccccl@{}}
\toprule
\textbf{Detector} & \textbf{Approach} & \textbf{Go} & \textbf{Java} & \textbf{JS/TS} & \textbf{Py} & \textbf{Targets} \\
\midrule
CodeQL & Static & \cmark & \cmark & \cmark & \cmark & CWE-tagged queries \\
Semgrep & Static & \cmark & \cmark & \cmark & \cmark & Route guards \\
BolaRay, BolaZ & BOLA & -- & -- & -- & -- & PHP only \\
BACFuzz & Dynamic & -- & -- & -- & -- & PHP only \\
Vulnhuntr & LLM & -- & -- & -- & \cmark & IDOR class \\
Claude Code & Agent & \cmark & \cmark & \cmark & \cmark & BAC prompt \\
Codex CLI & Agent & \cmark & \cmark & \cmark & \cmark & BAC prompt \\
Direct audit & Same LLM & \cmark & \cmark & \cmark & \cmark & BAC prompt \\
Trusting judge & Same LLM & \cmark & \cmark & \cmark & \cmark & Certificate \\
\bottomrule
\end{tabular}
\end{table}

\para{Baselines.}  \Cref{tab:baselines} lists baselines from each family of \cref{sec:intro}.  CodeQL~\cite{codeql}
runs its queries tagged CWE-284, 285, 639, 862, or 863 on the JS/TS, Python, and Java versions; none concerns
object-level authorization (its IDOR query~\cite{codeqlidor} is for C\#, and its one Go query models a PAM library that
none of our Go trees imports).  Semgrep~\cite{semgrep} runs the OWASP~A01 rules and route-detect~\cite{routedetect}.  BolaRay~\cite{huang2024bolaray}, BolaZ~\cite{bolaz}, and BACFuzz~\cite{dharmaadi2025bacfuzz} analyze PHP.  The LLM-based baselines use \sys's model.  Vulnhuntr~\cite{vulnhuntr}
analyzes the Python files at \sys's reasoning effort and output limit and flags an IDOR confidence of at least 7.
Claude Code~\cite{claudecode} and the Codex CLI~\cite{codexcli}, with read-only tools and default settings, are asked
for the endpoint's BAC vulnerabilities in the checked-out repository.
Two baselines hold everything but the verdict fixed.  The \emph{direct audit} gives the same model the same slice and asks
for the endpoint's object-level authorization flaws, and the \emph{trusting judge} reads \sys's certificate but
believes it---its coverage list and every grounded claim whose subject does not look like request data.

\para{Metrics.}  A detector \emph{flags} a version if it reports a finding inside the endpoint.  On \textsc{Pairs}, we
count the vulnerable versions flagged, the fixed versions flagged, and the pairs \emph{separated} (vulnerable flagged,
fixed not), averaged per repository (macro), because one repository contributes 15 of the 69 pairs.  On \textsc{Hunt}, a
finding has the \emph{right reason} when it is a definite finding at the access that the exploit abuses.

\para{Protocol.}  RQ2--RQ5 report the final \sys, with one certificate per endpoint and one objection round, on two
2.1-GHz Xeon cores with 8\,GB of memory.

\subsection{RQ1: Zero-Day Vulnerabilities}
\label{sec:rq1}

\begin{table}[!htb]
\caption{\RQtag{1} A selective list of zero-day vulnerabilities detected by \sys, and a breakdown of all 112 by the
clause of \cref{def:property} that they violate and by operation (R/W/D/C: read, write, delete, create).}
\label{tab:zeroday}
\centering
\footnotesize
\setlength{\tabcolsep}{2.6pt}
\begin{tabular}{@{}lllrrll@{}}
\toprule
\textbf{Application} & \textbf{Lang.} & \textbf{Kind} & \textbf{Stars} & \textbf{\#} & \textbf{Op.} & \textbf{Status} \\
\midrule
Gitea & Go & Code hosting & 59K & 1 & R & Reported \\
Grafana & Go/TS & Observability & 78K & 1 & W & Reported \\
Strapi & JS/TS & CMS & 73K & 12 & R/W/D & Reported \\
Plane & Python & Issue tracker & 62K & 6 & R/W/D & Reported \\
JumpServer & Python & Bastion host & 32K & 10 & R/W & Reported \\
Mattermost & Go & Boards plugin & 0.1K & 1 & W & Confirmed \\
\midrule
\multicolumn{7}{@{}l}{\emph{By clause}, wrong object 41; no check 40; client-chosen subject 12;} \\
\multicolumn{7}{@{}l}{\quad not enforced 9; weak level 5; uncovered key 4; written reference 1} \\
\multicolumn{7}{@{}l}{\emph{By operation}, read 82; write 17; delete 9; create 4} \\
\midrule
\textbf{Total} & & & & \textbf{112} & & \\
\bottomrule
\end{tabular}
\end{table}

We consider a finding a \emph{zero-day} if no public report of it existed and we confirmed it with a working exploit on
a local instance of the application, with two accounts of the lowest privilege that the endpoint admits.  \sys
detects 112 such vulnerabilities in 44 of the 57 applications, 74 of them in 25 projects with over 20,000 GitHub
stars.  \Cref{tab:zeroday} shows a selective list.  We have disclosed all of them
to the developers; Mattermost, for example, has confirmed our report and is assigning a CVE identifier.

The commonest flaws are a check on the wrong object (41) and no check (40); most are reads (82).  \sys reports 109 as definite
findings and two as leads; Mattermost's comes from the sibling norm, as undelete, unlike delete and patch, never asks
whether the caller owns the comment.  The vulnerabilities share one trait with \cref{lst:example}.  In most of them, the
endpoint does check \emph{something}---a session, a role, a parent, a predicate---and the flaw lies in what the check is
about, which is precisely what the conditions of \cref{tab:spec} decide.

\subsection{RQ2: Comparison with Existing Detectors}
\label{sec:rq2}

\begin{table}[!htb]
\caption{\RQtag{2} Known vulnerabilities and their fixes.  \textsc{Pairs} gives macro averages in percent over 35
repositories, and \textsc{Hunt} the endpoints flagged of 41 and, of these, those flagged for the right reason (for a baseline, a
location within five lines of the abused access).  Vulnhuntr runs on the Python versions
(8 pairs, 11 \textsc{Hunt} cases), where \sys flags 100 and 46.7\% and 9 of 11.}
\label{tab:compare}
\centering
\footnotesize
\setlength{\tabcolsep}{3pt}
\begin{tabular}{@{}lccccc@{}}
\toprule
& \multicolumn{3}{c}{\textbf{\textsc{Pairs} (macro \%)}} & \multicolumn{2}{c}{\textbf{\textsc{Hunt} (41)}} \\
\cmidrule(lr){2-4}\cmidrule(l){5-6}
\textbf{Detector} & \textbf{Vuln.}$\uparrow$ & \textbf{Fixed}$\downarrow$ & \textbf{Sep.}$\uparrow$ & \textbf{Flagged} & \textbf{Right} \\
\midrule
CodeQL (no Go) & 0.0 & 0.0 & 0.0 & 0 & 0 \\
Semgrep & 17.1 & 16.2 & 1.0 & 3 & 0 \\
Vulnhuntr (Python) & 60.0 & 60.0 & 20.0 & 7/11 & -- \\
Claude Code (agent) & 91.9 & 12.3 & \textbf{79.6} & 34 & 21 \\
Codex CLI (agent) & \textbf{97.1} & 22.9 & 74.3 & \textbf{40} & 23 \\
Direct audit & 46.7 & 11.1 & 36.7 & 26 & -- \\
Trusting judge & 71.0 & 36.9 & 38.8 & 28 & -- \\
\midrule
\sys, tier A & 47.9 & \textbf{4.5} & 46.2 & 19 & 17 \\
\sys, tiers A--B & 55.0 & 9.7 & 51.0 & 24 & 20 \\
\sys, all definite & 85.2 & 27.2 & 62.8 & 33 & \textbf{28} \\
\sys, with leads & 86.2 & 32.6 & 57.4 & 34 & -- \\
\bottomrule
\end{tabular}
\end{table}

\begin{figure}[!t]
  \centering
  \includegraphics[width=\columnwidth]{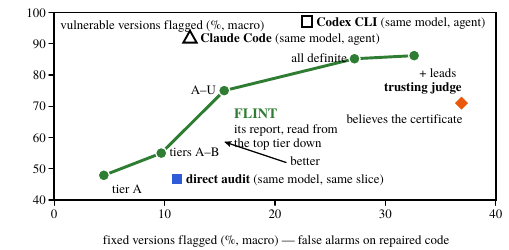}
  \caption{\RQtag{2} Vulnerable versions flagged (higher is better) against fixed versions flagged, i.e., false alarms on
  repaired code (lower is better), macro over \textsc{Pairs}.  Each point of \sys's curve reads its report down to one more
  tier; at every depth, it lies above and left of (better than) the same model given the same slice, auditing it directly
  or believing \sys's certificate.}
  \label{fig:pairs}
\end{figure}

\para{Known Vulnerabilities and Their Fixes.}  \Cref{tab:compare} and \cref{fig:pairs} show the results.  The static
analyzers report almost nothing at the flaw, and Semgrep's route rules fire on vulnerable and fixed versions alike.  The
coding agents, driven by the same model over the whole repository, are the strongest baselines: they separate 79.6 and
74.3\% of the pairs against \sys's 62.8\%.  The gap lies mostly on repaired code, where \sys keeps an access whose new guard it cannot verify and an
agent judges the guard sufficient.  Claude Code takes 0.19\,CNY and a median of 167\,s per endpoint, four to five times
\sys's cost and model time.  On \textsc{Hunt}, the agents flag more endpoints but fewer for the right reason (21 and 23
against 28), and both find \cref{lst:example}.  The direct audit flags fewer than half of the vulnerable versions; the
trusting judge flags more, but fixed versions more than three times as often, so believing the certificate does not
separate vulnerable from repaired code.  Verification does: read from the top tier down, \sys's report dominates the
direct audit at every depth.  Tier~A alone matches its recall (47.9 against 46.7\%) with less than half of its false
alarms, and tiers A--B improve all three measures.  The gain is largest on Java, where the direct audit flags 4 of 22
vulnerable versions and \sys 18 (15 pairs separated against 4), and smallest on JS/TS, where both separate 5 of 13.

On \textsc{Hunt}, every endpoint that the direct audit flags, \sys flags too, and seven only \sys flags, among them
\cref{lst:example} and the case study below.  Counting every handler that reaches a flaw located in a service, \sys reports 38 of the 43
object-level vulnerabilities, 31 for the right reason.  Of the five it misses, two are accesses that the certificate does
not list (a reference written by mass assignment, an order-code lookup), two lie outside every analyzed function, and one
record is unusable.

\para{Near Misses.}  A check that looks like the fix but does not restrict the access tests whether a detector verifies
a guard or trusts it.  \textsc{Near} mutates each expressible fix of \textsc{Pairs} with one operator that violates a
clause of \cref{def:property}, so that the original attack works again: a weaker permission (20 fixes), a subject that
is not the caller (4), or a failure that is logged and ignored (9); two fixes admit no such one-line mutation.  Of the 31
near misses with every detector's answer, the direct audit flags 14, \sys 24 (26 with leads), Claude Code 27, and the
Codex CLI 30 (McNemar against \sys, $p = 0.45$ and $0.07$); \sys clears the real fix and flags its near miss in 18 pairs,
the agents in 24 and 23.  \sys misses three near misses whose certificate lists no access, two that remain leads, and two
cleared by a permission whose name does not say read, which the challenge does not question.  On the 53 confirmed vulnerabilities whose report
names a check that leaves the access open (\textsc{Hunt} and RQ1), \sys finds the abused access in 18 of the 32 with a
known access, Claude Code in 12, and the Codex CLI in 15 (\cref{tab:near}).

\para{Next to a Vulnerability.}  On \textsc{Neigh}, \sys reports 33 of the 41 vulnerable endpoints (80\%) and 118 of
the 227 siblings (52\%) in a definite tier; the direct audit flags 26 (63\%) and 59 (26\%).  Of the 37 siblings in
\sys's top tier, all of which we read, 24 repeat the known flaw of their file, 8 are real but minor, and 5 are not findings.  Of the 9
siblings that the direct audit flags and \sys clears, 6 are false alarms of the direct audit, which missed a guard on the
declaration or in a service, 2 are misses of \sys, and 1 record is unusable.

\para{Case Study.}  One vulnerability that \sys reports and the direct audit misses lets a staff member who may manage
customers take over administrator accounts of a commerce platform: the customer-update mutation checks the
customer-management permission but resolves its target among all users, so the caller can rewrite an administrator's
e-mail address and reset the password.  The model waived the finding as \kind{NOT\_NEEDED}, but the permission is a role
gate that concerns no object (V4), so the finding stayed in the report, in tier~C.

\subsection{RQ3: Ablation Study}
\label{sec:rq3}

\begin{table}[!htb]
\caption{\RQtag{3} Ablation study, removing one component at a time.  Siblings cleared on \textsc{Neigh} measure how many
guards the verifier accepts (stage one).  A dash marks a measure that does not apply or that we did not run.}
\label{tab:ablation}
\centering
\footnotesize
\setlength{\tabcolsep}{2.5pt}
\begin{tabular}{@{}lcccc@{}}
\toprule
& \multicolumn{2}{c}{\textbf{\textsc{Pairs} (macro \%)}} & \textbf{\textsc{Hunt}} & \textbf{\textsc{Neigh}} \\
\cmidrule(lr){2-3}
\textbf{Configuration} & \textbf{Vuln./Fixed} & \textbf{Sep.} & \textbf{flagged} & \textbf{sib.\ cleared} \\
\midrule
\sys & 85.2 / 27.2 & 62.8 & 33 & 94 \\
$-$ authority analysis & 72.5 / 17.1 & 61.1 & 26 & 95 \\
$-$ verification (trusting judge) & 71.0 / 36.9 & 38.8 & 28 & -- \\
$-$ certificate (direct audit) & 46.7 / 11.1 & 36.7 & 26 & -- \\
$-$ objection round & 85.2 / 30.8 & 59.2 & 34 & -- \\
$-$ need challenge & 85.2 / 27.2 & 62.8 & 32 & 95 \\
$-$ attestation (V7) & 85.2 / 27.2 & 62.8 & 32 & -- \\
$-$ completeness rules (each) & -- & -- & 33--35 & 81--94 \\
$-$ verifier-driven demand & 84.5 / 27.9 & 62.1 & 33 & 91 \\
Other model (V4-Pro) & 62.0 / 40.7 & 29.6 & 32 & -- \\
\bottomrule
\end{tabular}
\end{table}

\para{Authority Analysis.}  \Cref{tab:ablation} removes one component at a time.  Without the dataflow analysis, a \VALID check counts wherever it stands.  \sys then flags 49
instead of 61 vulnerable versions of \textsc{Pairs} and 26 instead of 33 on \textsc{Hunt}.  Twelve vulnerable versions
and seven known vulnerabilities are thus definite findings only because an existing check does not stand on the path of the
access---after it, on another branch, or behind a client-selected alternative.  On \textsc{Neigh}, seven
vulnerable endpoints lose their definite finding.

\para{Stage Two and Another Model.}  The sibling norm, which has no row in \cref{tab:ablation}, adds one of the 41
\textsc{Neigh} vulnerabilities and lifts three more and one sibling into tier~A*, but none of the 20 new vulnerabilities of
RQ1 that stage one misses.  With DeepSeek-V4-Pro for every neural step, \sys separates 29.6\% of the \textsc{Pairs}; the
larger model lists no access in 22 of 141 certificates (V4.1-Flash in 6) and claims fewer checks
(2.7 against 3.1), and our prompts were written with V4.1-Flash.

\para{The Loop.}  The objection round cleared three fixed versions with claims that verified (a scope by workspace and
the owner comparison that a fix added) and no vulnerable version.  The verifier demanded code for 63\% of the
\textsc{Pairs} endpoints with an objection (538 function bodies) and 75\% of the \textsc{Hunt} ones; without it, 91 instead
of 94 siblings are cleared, and tier~A holds 25 instead of 29 vulnerable versions of \textsc{Pairs}.  The need challenge
questioned 4, 2, and 7 accesses of \textsc{Pairs}, \textsc{Hunt}, and \textsc{Neigh}, and the majority named a finer
required permission for none, 2, and 3, which adds one \textsc{Hunt} endpoint, one sibling, and five near misses.
Thirteen rules (\cref{tab:rules}) make verification complete enough to accept real guards: switched off one at a time,
ten of them each leave two to thirteen more siblings uncleared, and the vulnerable endpoints that come back are flagged
for a wrong reason.  One rule, V7, adds a detection: without it, \cref{lst:example} is cleared on the model's defense.

\subsection{RQ4: Soundness of Clearance}
\label{sec:rq4}

\begin{table}[!htb]
\caption{\RQtag{4} Fabricated (Fabr.) claims, in the certificate (cert.) or as a defense, and adversarial samples (34 samples, 50,903 claims).  A claim is accepted if it
verifies as an object check, not merely as a gate; an endpoint is cleared if its finding is gone.}
\label{tab:attacks}
\centering
\footnotesize
\setlength{\tabcolsep}{2.4pt}
\begin{tabular}{@{}llrrrr@{}}
\toprule
\textbf{Attack} & \textbf{Set (flagged)} & \textbf{Claims} & \textbf{Accepted} & \textbf{Cleared} & \textbf{To lead} \\
\midrule
Fabr.\ cert. & \textsc{Pairs} (63) & 946 & 92 & 1$^\dagger$ & 32 \\
& \textsc{Hunt} (33) & 517 & 71 & 0 & 20 \\
\cmidrule(lr){1-6}
Fabr.\ defense & \textsc{Pairs} (63) & 946 & 92 & 1$^\dagger$ & 0 \\
& \textsc{Hunt} (33) & 517 & 71 & 0 & 0 \\
\midrule
Adversarial & \textsc{Hunt} (33) & 443--547 & 145--216 & 1$^\ddagger$ & 0 \\
& \textsc{Pairs} vuln.\ (63) & 558--905 & 171--279 & 0--1$^\ddagger$ & 0 \\
& \textsc{Pairs} fixed (23) & 189--234 & 68--101 & 1--3 & 0--1 \\
\bottomrule
\multicolumn{6}{@{}p{0.97\columnwidth}@{}}{\scriptsize $^\dagger$One endpoint, flagged for a wrong reason; the claims are
true.  $^\ddagger$The same endpoint in every sample; its flagged access is guarded, and the flaw lies elsewhere.}
\end{tabular}
\end{table}

\Cref{lem:inert} says that claims that do not verify cannot clear; a claim that \emph{does} verify is the residual
risk, which we measure with claims that are false by construction and with a model that tries.

\para{Fabricated Claims.}  We give every uncleared access of a flagged endpoint claims that cite real code but are false.
Ten fault models present a call, a branch, the fetch, a role gate, or the access itself as a check on the flagged
object, or re-claim an existing check for it.  Each claim asks for the highest level, once in the certificate and once in defense of an objection.  Of
2,926 claims (\cref{tab:attacks}), 558 relabel a role gate and verify as a gate, which covers no object; 326 more verify,
almost all of them true statements about code that does not stand on the access.  Four claims cleared an access, all on one endpoint flagged for a wrong reason, where they happen to be true.
\emph{No fabricated claim cleared an access known to be unguarded}; in defense, unverifiable claims changed nothing.

\begin{figure}[!t]
  \centering
  \includegraphics[width=\columnwidth]{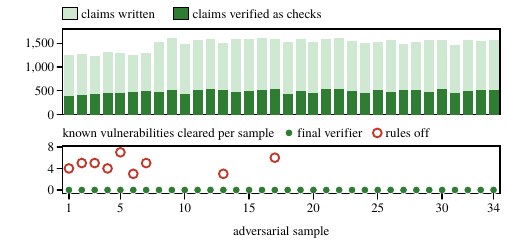}
  \caption{\RQtag{4} Adversarial samples.  Top, the claims written by a model told to have every objected access
  accepted and those that verified as checks; bottom, the known vulnerabilities cleared per sample with the final
  verifier and with the 34 hardening rules off (nine stored samples).}
  \label{fig:adv}
\end{figure}

\para{Adversarial Samples.}  A fabricated claim is random; a model that tries is not.  We showed the model the
verifier's objections and told it to make the verifier accept every objected access, with whatever real code could pass
for a restriction.  In 34 samples over \textsc{Pairs} and \textsc{Hunt}, up to 279 claims per sample verify, yet none clears or
demotes an access known to be unguarded; the vulnerable endpoints that do clear are guarded where stage one flags them,
and their flaw lies elsewhere.  Only the need challenge rests on the model's word; a model that tries would withhold the
one \textsc{Hunt} endpoint that it adds.  The first sample, drawn before any hardening, cleared three of the 34 known
vulnerabilities that it attacked, and without the 34 hardening rules, nine stored samples clear three to seven each
(\cref{fig:adv}).  Each hole that a sample found became a rule; of the 27 samples drawn after the rules
from probing, three found one, the last against the verifier that we had declared final.  Further samples can find
further holes, in a claim field or an assumption (A1--A5) that verification takes on trust.

\subsection{RQ5: Practicality}
\label{sec:rq5}

\begin{figure}[!t]
  \centering
  \includegraphics[width=\columnwidth]{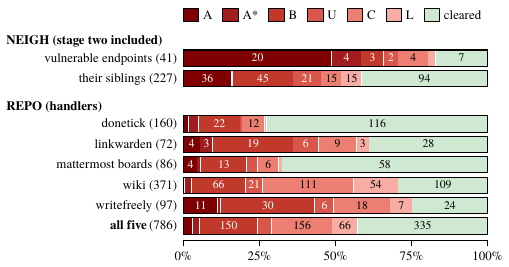}
  \caption{\RQtag{5} The report by tier on \textsc{Neigh} and on five whole applications (786 handlers): A, conceded with
  an attack; A*, below the siblings' verified norm; B, refuted defense; U, undecided; C, waived; L, lead.}
  \label{fig:tiers}
\end{figure}

\para{The Report.}  On the five applications of \textsc{Repo}, \sys clears 335 of the 786 handlers (42.6\%) and puts 42
(5.3\%) in the top tier (\cref{fig:tiers}); 385 (49.0\%) carry a definite finding in some tier, so the
order is what makes the report usable.  All five known vulnerabilities of these applications are reported, three in the
top tier and two in tier~B.  We read all 24 tier-A entries.  Of these, 8 are real or plausible (2 of them known), 7 have
low impact (existence oracles, a cache overwrite, a forced side effect), and 9 are not findings (5 by design, 2 where the
need challenge named a permission that the application does not require, one websocket callback, one documented public
endpoint).  Tier~C is dominated by one application whose public read endpoints the model declares unrestricted by design
(111 of its 156 entries).

\begin{figure}[!t]
  \centering
  \includegraphics[width=\columnwidth]{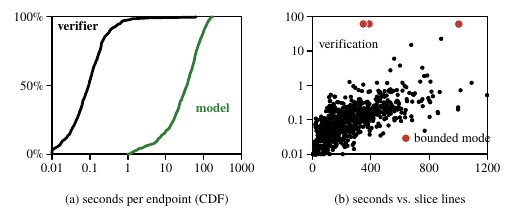}
  \caption{\RQtag{5} Time per endpoint on \textsc{Repo} (831 functions): (a) model time (certificate and need round) and
  verification time; (b) verification time against slice size, with the three bounded endpoints at the time limit.}
  \label{fig:time}
\end{figure}

\para{Time and Cost.}  \Cref{fig:time} shows where the time goes.  The certificate of an endpoint, need round included,
takes a median of 36\,s of model time (90th percentile 100\,s), and calls run concurrently.  Verification needs no
model.  The 831 functions of \textsc{Repo} take 6.0 minutes on one core, with a median of 0.09\,s and a 99th percentile of
1.9\,s per endpoint; verification time grows with the slice but is dominated by origin tracing on a few endpoints, three
of which exceed the time limit and run in the bounded mode.  A certificate takes 9.3\,K prompt and 8.2\,K
completion tokens on average, US\$0.006 per endpoint at off-peak prices.

\para{Completeness of a Clearance.}  A cleared endpoint is cleared on the accesses that its certificate lists.  A
model-free check finds that 108 of the 335 cleared handlers of \textsc{Repo} call repository code outside their slice;
in 70 the call is itself a listed access, and in 28 the callee touches none of the data-access objects that the
application's certificates use.  The remaining 10 (3\%) may hide an access that no certificate lists.  The check
marks them as leads; it was added after the reported run, so \cref{fig:tiers} counts them as cleared.

%% file: sections/discussion.tex
\section{Discussion}
\label{sec:discussion}

\para{Limitations.}  \Cref{thm:clearance} speaks about the accesses that the certificate lists.  An access that the
model omits is never judged, e.g., references written by mass assignment~\cite{corradini2023mass} (\cref{sec:rq2}), and
3\% of the cleared handlers of whole applications call code outside the slice that may access data (\cref{sec:rq5}).
\sys does not verify recall; it ensures that an omission never comes with a clearance that reads as a proof.  The guarantee also takes A3 and A4 on trust, as both need the application's
policy (\cref{sec:loop}).

\para{Non-Determinism and Untrusted Code.}  Verification is deterministic given a certificate; certificates are not.  A
second certificate for every \textsc{Pairs} version flags 61 vulnerable and 34 fixed versions, against 61 and 21 for the
first; requiring both to agree gives 81.0/23.0/63.2\% (macro) instead of 85.2/27.2/62.8\%.  Attacker-written text that steers the model~\cite{greshake2023indirect} can make it omit
an access, which lowers recall, or invent a check, which cannot clear anything (\cref{lem:inert}).

%% file: sections/related.tex
\section{Related Work}
\label{sec:related}

\para{Access-Control Flaws in Web Applications.}  Static detectors infer what a check should be from the code.
RoleCast~\cite{son2011rolecast} and Sun \etal~\cite{sun2011static} compare roles, MACE~\cite{monshizadeh2014mace}
checks a query's authorization context across paths, SPACE~\cite{near2016space} matches Rails patterns, and
FixMeUp~\cite{son2013fixmeup} repairs missing checks.  BolaRay~\cite{huang2024bolaray} and BolaZ~\cite{bolaz} derive object-level authorization
models from the SQL of PHP applications, and MOCGuard~\cite{liu2025mocguard} finds missing owner checks across Java and
SQL; on mobile back ends, multi-step authorization flaws are exploited by chaining requests~\cite{shi2026foot}.  These detectors are precise where their model of the data layer holds; \sys needs no such model, so one
implementation covers five languages, and in exchange it inherits the LLM's recall.  Runtime and
black-box approaches~\cite{dalton2009nemesis,felmetsger2010logic,pellegrino2014logic,doupe2012enemy,dharmaadi2025bacfuzz,
rennhard2022automating,autorize,authmatrix,atlidakis2020checking} cover only what a running
deployment lets them reach~\cite{doupe2010johnny,bau2010state,eriksson2021blackwidow,atlidakis2019restler}, and
language-based approaches~\cite{chlipala2010urflow,near2012rubicon,zhang2002cqual} need applications written for them.

\para{Missing Checks and Static Analysis.}  Missing-check detectors for systems code infer the norm from the
code~\cite{engler2001deviant,tan2008autoises,li2005prminer,yamaguchi2013chucky,shao2016kratos,zhang2019pex,lu2019crix,
lu2022mpchecker,yun2016apisan}; \sys's sibling norm applies this idea, but a norm counts only where verification
confirmed it.  Taint and dataflow analyses find injection and pollution flaws in web code~\cite{livshits2005finding,
jovanovic2006pixy,balzarotti2007multimodule,son2011saferphp,li2022odgen,kang2022probe,kang2025gala,liu2025domino,liu2026pyrl}, also as
queries~\cite{yamaguchi2014cpg,avgustinov2016ql,codeql,semgrep,li2014survey}; for authorization, origin tracing must
tell \emph{whose} value an operand is, not only whether it is tainted.

\para{LLMs and Program Analysis.}  LLMs propose what classical analyses need, e.g., taint
specifications~\cite{li2025iris}, dataflow facts~\cite{wang2024llmdfa,wang2024llmsa}, bug-specific checks and
queries~\cite{li2024llift,yang2025knighter,li2026mocq}, and slices to classify~\cite{lekssays2025llmxcpg}; auditors and
agents read whole repositories~\cite{guo2025repoaudit,vulnhuntr,semgrepai,sun2024gptscan,bigsleep2024,
claudecodesecurity,codexsecurity}.  Most let the model end the analysis, which studies find
fragile~\cite{ullah2024llms,ding2025primevul,steenhoek2024err}.  In verification, LLMs propose invariants, proofs, and
properties for a prover~\cite{wu2024lemur,chakraborty2023ranking,first2023baldur,liu2025propertygpt}; \sys's prover
checks only the model's defenses, in the spirit of proof-carrying code~\cite{necula1997pcc}, certifying
algorithms~\cite{mcconnell2011certifying}, and translation validation~\cite{pnueli1998translation}.

%% file: sections/conclusion.tex
\section{Conclusion}
\label{sec:conclusion}

We presented \sys, which detects broken access control by separating the proposal from the verdict.  An LLM proposes,
per endpoint, a certificate of the data accesses and the checks that guard them, and a symbolic verifier decides
every access; a check clears an access only if the code
shows that it binds the caller to the selected object on every path.  On known vulnerabilities and their fixes,
\sys's report dominates the same model auditing the code directly, and on popular applications, \sys found 112 zero-day vulnerabilities.

%% file: sections/appendix.tex
\section{Exploits for the Motivating Example}
\label{app:exploit}

\Cref{tab:exploit} shows the requests of the exploit against \cref{lst:example}, sent to a local instance of the
application.  Users A and B are members of the same organization with the lowest default role; A is the attacker.  B first creates a personal access token, whose identifier is \code{<B-token>}.  In the table, \code{P} stands for the path
\code{/api/v1/user/me/personal-access-tokens}, and B$^{*}$ is a request authenticated with the secret of B's token.  The
two control requests show that the
application otherwise denies A access to B's token.  That is, its sibling endpoint, which rotates a token, reads it through a query
scoped by the caller and answers 404.  The attack deletes B's token, and B's credential stops authenticating at once.

\begin{table}[h]
\caption{Requests of the exploit against \cref{lst:example}.}
\label{tab:exploit}
\centering
\footnotesize
\setlength{\tabcolsep}{3.5pt}
\begin{tabular}{@{}lll@{\hspace{1em}}l@{}}
\toprule
\textbf{Step} & \textbf{As} & \textbf{Request} & \textbf{Response} \\
\midrule
Setup & B & \tcode{POST P} & 200, \tcode{<B-token>} \\
Control & A & \tcode{PATCH P/<B-token>/rotate} & 404 \\
Control & A & \tcode{DELETE P/<A-token>} & 200 \\
\textbf{Attack} & A & \tcode{DELETE P/<B-token>} & \textbf{200} \\
Impact & B$^{*}$ & \tcode{GET /api/v1/user} & 401 \\
Impact & B & \tcode{GET P} & 200, \tcode{[]} \\
\bottomrule
\end{tabular}
\end{table}

\section{Near Misses and Natural Decoys}
\label{app:near}

\Cref{tab:near} breaks the near misses of \cref{sec:rq2} down by operator and adds the natural decoys, i.e., the
confirmed vulnerabilities of \textsc{Hunt} and RQ1 whose report names a check that leaves the access open, next to those
with no check beyond authentication.  The agents flag more near misses with a weaker permission, where the permission
that the operation requires is finer than read and write (\cref{sec:discussion}).  On natural decoys, \sys finds the
abused access more often than either agent, but no difference is significant (exact McNemar, $p > 0.1$).

\begin{table}[h]
\caption{Near misses of \textsc{Near} (31 fixes with every detector's answer) by operator, pairs told apart (fix
cleared, near miss flagged), and natural decoys found for the right reason (confirmed vulnerabilities whose abused
access is known, with a check that leaves it open or with no check beyond authentication).}
\label{tab:near}
\centering
\footnotesize
\setlength{\tabcolsep}{3pt}
\begin{tabular}{@{}lrrrrr@{}}
\toprule
& \textbf{\sys} & \textbf{+leads} & \textbf{Direct} & \textbf{Claude} & \textbf{Codex} \\
\midrule
\multicolumn{6}{@{}l}{\textsc{Near}} \\
Vulnerable flagged & 27 & 27 & 13 & 29 & 30 \\
Real fix flagged & 7 & 8 & 4 & 3 & 7 \\
Near miss: weaker level (19) & 13 & 15 & 4 & 15 & 18 \\
Near miss: other subject (4) & 4 & 4 & 3 & 4 & 4 \\
Near miss: not enforced (8) & 7 & 7 & 7 & 8 & 8 \\
Pair told apart & 18 & 19 & 10 & 24 & 23 \\
\midrule
\multicolumn{6}{@{}l}{Natural decoys, right reason} \\
Decoy check (32) & 18 & -- & -- & 12 & 15 \\
No check (43) & 27 & -- & -- & 22 & 25 \\
\bottomrule
\end{tabular}
\end{table}

\section{Completeness Rules}
\label{app:rules}

\Cref{tab:rules} lists the thirteen rules of \cref{sec:rq3}, each switched off alone.  Without ten of them, more siblings
on \textsc{Neigh} stay uncleared, and the \textsc{Hunt} endpoints that come back are flagged for a wrong reason: the
count flagged for the right reason never rises.  Without V7, \cref{lst:example} is cleared.

\begin{table}[h]
\caption{The thirteen rules of \cref{sec:rq3}, each switched off alone: \textsc{Hunt} endpoints flagged (for the right
reason) and \textsc{Neigh} siblings cleared at stage one.}
\label{tab:rules}
\centering
\footnotesize
\setlength{\tabcolsep}{3pt}
\begin{tabular}{@{}p{0.68\columnwidth}cc@{}}
\toprule
\textbf{Rule switched off} & \textbf{\textsc{Hunt}} & \textbf{Sib.} \\
\midrule
None (\sys) & 33 (28) & 94 \\
\midrule
The request parameter of a guard that the framework invokes is the handler's request & 35 (28) & 83 \\
A guard that the framework invokes binds to the object of the same request key & 33 (28) & 91 \\
A check on what an object refers to (by a field of its fetched row) covers the object & 33 (28) & 92 \\
Origins flow back through the return values of slice functions (\rn{Ret}) & 34 (28) & 88 \\
A field of a nested literal among a call's arguments is no field of the result & 33 (28) & 92 \\
Guards that the declaration, class header, or route registration names apply & 35 (28) & 83 \\
\quad and so do guards that an applied guard names & 35 (28) & 81 \\
Code asked for in the need round is added inside files of the slice too & 33 (28) & 88 \\
The operands of one lookup are joint, whatever variable they come from & 33 (28) & 92 \\
A field write to a parameter reaches only the uses that follow it & 33 (28) & 90 \\
V7: a predicate handed a record assembled from the caller's identity & 32 (27) & 94 \\
V2 also for links; a call that computes a stored value is no test & 33 (28) & 94 \\
Functions handed to the router before the route apply to it & 33 (28) & 94 \\
\bottomrule
\end{tabular}
\end{table}

\section{Use of Generative AI}
\label{app:genai}

An LLM is a component of \sys.  The model, its settings, and how its answers are cached are given in \cref{sec:impl},
and every baseline that uses a model is described in \cref{sec:setup}.  %
Generative AI was used for editorial purposes in this manuscript.